\documentclass[11pt]{article}
\usepackage{amsfonts}
\usepackage{amsfonts}
\usepackage{amsfonts,amsmath,amssymb,epsf}
\usepackage{graphicx,color}
\usepackage[usenames,dvipsnames]{pstricks}
\usepackage{epsfig}
\usepackage{pst-grad} 
\usepackage{pst-plot} 
\usepackage{verbatim}

\newcommand{\al}{\alpha'}
\newcommand{\de}{\partial}
\newcommand{\be}{\begin{equation}}
\newcommand{\ba}{\begin{eqnarray}}
\newcommand{\ea}{\end{eqnarray}}
\newcommand{\ee}{\end{equation}}

\newcommand{\we}{\wedge}

\newcommand{\f}{\frac}
\newcommand{\s}{\sqrt}
\newcommand{\vp}{\varphi}

\newcommand{\ti}{\tilde}
\newcommand{\ap}{\alpha}

\newcommand{\ddd}{\cdot\cdot\cdot}
\newcommand{\no}{\nonumber \\}

\newcommand{\ep}{\epsilon}

\newcommand{\x}{\times}
\begin{document}

\begin{titlepage}
\thispagestyle{empty}

\begin{flushright}
KUNS-2177

IPMU09-0002
\end{flushright}

\bigskip

\begin{center}
\noindent{\Large \textbf{Fractional Quantum Hall Effect via Holography:\\
Chern-Simons, Edge States, and Hierarchy}}\\
\vspace{15mm} Mitsutoshi Fujita$^{a}$\footnote{e-mail:
mfujita@gauge.scphys.kyoto-u.ac.jp},\ Wei Li$^{b}$\footnote{e-mail:
wei.li@ipmu.jp},\ Shinsei Ryu$^{c}$\footnote{e-mail:
sryu@berkeley.edu} and
Tadashi Takayanagi$^{b}$\footnote{e-mail: tadashi.takayanagi@ipmu.jp}\\
\vspace{1cm}

 {\it $^{a}$Department of Physics, Kyoto University, Kyoto 606-8502, Japan \\
 $^{b}$Institute for Physics and Mathematics of the Universe (IPMU), \\
 University of Tokyo, Kashiwa, Chiba 277-8582, Japan\\
  $^{c}$Department of Physics, University of California, Berkeley, CA 94720, USA}

\vskip 3em
\end{center}

\begin{abstract}

We present three holographic constructions of fractional quantum
Hall effect (FQHE) via string theory. The first model studies edge
states in FQHE using supersymmetric domain walls in ${\cal N}=6$
Chern-Simons theory. We show that D4-branes wrapped on
$\mathbb{CP}^1$ or D8-branes wrapped on $\mathbb{CP}^3$ create edge
states that shift the rank or the level of the gauge group,
respectively. These holographic edge states correctly reproduce the
Hall conductivity. The second model presents a holographic dual to
the pure $U(N)_k$ (Yang-Mills-)Chern-Simons theory based on a D3-D7
system. Its holography is equivalent to the level-rank duality,
which enables us to compute the Hall conductivity and the
topological entanglement entropy. The third model introduces the
first string theory embedding of hierarchical FQHEs, using IIA
string on $\mathbb{C}^2/Z_{n}$.
\end{abstract}

\end{titlepage}

\newpage

\section{Introduction and Summary}

The fractional quantum Hall effect (FQHE) is fascinating not only
because its extraordinary experimental realizations, but more
importantly, because it is a manifestation of the fundamental
concept of topological phase (see
e.g.\,\cite{PrangeGirvin,Wen,Yoshioka,Zee}). The fractional quantum
Hall states are characterized by the filling fraction
$\nu$.\footnote{The filling fraction $\nu$ is defined as
$\nu=\frac{N_e}{N_{\phi}}$, where $N_e$ is the total electron number
and $N_{\phi}$ the total flux number (which in turn equals the
maximal number of electrons that can be accommodated in each Landau
level).} When $\nu$ is an integer, it is called the integer quantum
Hall effect.

The simplest series of the fractional quantum Hall states is known
as the Laughlin state which has $\nu=\frac{1}{k}$ with $k$ being an
odd integer\footnote{
 Odd integer values of $k$ are realized when the elementary particles
 are
fermions (i.e. electrons), for which the Laughlin state was constructed. Even integer values of $k$ also occur in realistic materials
such as rotating cold atoms, as we will mention later.
In this paper, we are mainly working with the effective Chern-Simons description 
of FQHE (which does not depend on the statistics of elementary particles), 
without asking much about microscopic origins of the effective field theory.
Therefore, we consider both cases of even and odd integer values of $k$ in our models.}.
At low energies, the Laughlin states are described by a
(2+1)-dimensional $U(1)$ Chern-Simons theory coupled to an external
electromagnetic field $\ti{A}$ \cite{Wen}: \be
S_{QHE}=\f{k}{4\pi}\int A\we dA+\f{e}{2\pi}\int \ti{A}\we dA ,
\label{csd} \ee where $A$ is the $U(1)$ gauge field that describes
the internal degrees of freedom and $e$ is the charge of the
electrons. It follows straightforwardly from (\ref{csd}) that the
Hall conductivity (defined by $j_x=\sigma_{xy}E_y$) is fractionally
quantized: $\sigma_{xy}=\f{\nu e^2}{h}$. Since the Chern-Simons
gauge theory is topological \cite{WiCS} and has no propagating
degree of freedom in contrast to the Maxwell theory, the FQHE
provides an example for the gapped many-body system whose ground
state can be described by a topological field theory.

A more generic category of FQHEs is
defined by filling fractions which are continued fractions: \be
\nu=\f{1}{a_1-\f{1}{a_2-\f{1}{a_3-\ddd}}}. \ee Such FQHE appears in
real materials and has a fascinating underlying structure called
hierarchy: they are described by a series of $U(1)$ Chern-Simons
gauge fields, each providing quasi-particles that act as constituent
particles that form the quasi-particles in the next level---hence
the name ``hierarchy "\cite{Yoshioka,PrangeGirvin,Wen,Zee}.

\begin{figure}[htbp]
   \begin{center}
     \includegraphics[height=5cm]{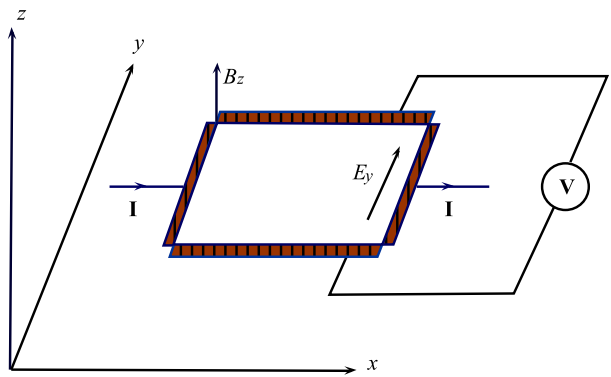}
   \end{center}
   \caption[Experimental setup of quantum Hall effect.]{Experimental
setup of the quantum Hall effect: a sample of a two-dimensional
electron gas is placed on the $xy$-plane, with a magnetic field
$B_z$ perpendicular to the plane and an electric field $E_y$ along
the $y$-direction. The quantum Hall current flows along the
$x$-direction, perpendicular to the $\vec{E}$-field.}
\label{fig:Experiment}
\end{figure}

Fig. \ref{fig:Experiment} shows the standard experimental
realization of QHE: a sample of two-dimensional electron gas is
placed on the $xy$-plane, with a magnetic field $B_z$ applied
perpendicular to the plane. The sample has four edges, to which four
probes are attached. To measure the Hall conductivity $\sigma_{xy}$,
we apply an electric field $E_y$ in the $y$-direction and measure
the current $j_x$.\footnote{Equivalently, we can drive current $j_x$
along the $x$-direction and measure the induced electric field
$E_y$.} Since there is an energy gap and realistic samples contain
impurities, the electrons in the (2+1)-dimensional bulk suffer from
localization and cannot move around macroscopically, thus the bulk
electrons do not contribute to the quantum Hall conductivity.
Instead, the electrons can only move along the edges of samples,
since the electron orbit at the boundary is stable against
impurities. These are called the edge states (for more details, see
textbooks \cite{Yoshioka,PrangeGirvin,Wen,Zee}). In other words, the
quantum Hall fluid is an insulator except at its boundaries. For
this reason, quantum Hall states are sometimes called ``topological
insulator". More general topologically insulating states, including
the quantum spin Hall state and its higher-dimensional relatives,
have been discussed, fully classified, \cite{topinsulators} and
experimentally realized recently \cite{topinsulators_exp}. In terms
of the Chern-Simons theory, the quantization of the Hall
conductivity can be understood as follows: since the Chern-Simons
action is not gauge invariant in the presence of (1+1)-dimensional
boundaries of the (2+1)-dimensional spacetime, there exists a
massless chiral scalar field on the boundaries \cite{Wen} and they
contribute to the conductivity.

The main purpose of this paper is to model fractional quantum Hall
effect---in particular its Chern-Simons theory description---in
string theory and analyze them via AdS/CFT
\cite{Maldacena}.\footnote{ Earlier models of embedding the FQHE in
string theory mainly used the noncommutative Chern-Simons
description (as opposed to holographic description)
\cite{Su,Polychronakos:2001mi,HeRa,HeSu,Br,BST,BOB}. Recently a holographic construction
of the QHE which is different from ours was realized in
\cite{KK,DKS} (for a related setup see \cite{Rey}). See also
\cite{HK,OB,MW} for holographic calculations of the classical Hall
effect.} We will present three different models, each focusing on
different aspects of the FQHE. They are summarized as follows.

\paragraph{Model I: Supersymmetric edge states in ${\cal N}=6$ Chern-Simons theory.} The first model is
based on the recently discovered $AdS_4/CFT_3$ correspondence:
${\cal N}=6$ Chern-Simons theory as the holographic dual to type IIA
string theory on $AdS_4 \times \mathbb{CP}^3$ \cite{ABJM}. We will
construct two types of supersymmetric D-brane configurations to
model the edge state of FQHE. First, we wrap $M$ D4-branes on the
$\mathbb{CP}^1$ inside the $\mathbb{CP}^3$ and attach them to a
stack of $N$ D2-branes along one spatial dimension; the
$(1+1)$-dimensional intersection acts as a domain wall across which
the gauge group $U(N)_{k}\times U(N)_{-k}$ jumps into
$U(N-M)_{k}\times U(N)_{-k}$. Similarly, we can also wrap $l$
D8-branes on the entire $\mathbb{CP}^3$ and attach them to the $N$
D2-branes in the same way; the gauge group now jumps into
$U(N)_{k-l}\times U(N)_{-k}$ upon crossing the edge.

These intersecting D-brane configurations with generic $(N,k)$ and
$M$ (or $l$) are interesting in their own right: first, they are new
examples of supersymmetric edge states in ${\cal N}=6$ Chern-Simons
theory; second, they are stable configurations that break the parity
symmetry.\footnote{This is consistent with the interpretation of
 fractional branes in \cite{ABJ}.} However, in
this paper, we will only focus on their application in modeling FQHE
holographically. To model a realistic FQHE, we simply set $N=1$ and
treat the $U(N)_{-k}$ part as inspector. Then the single D4-brane
intersection produces a standard edge state for FQHE with
$\nu=\frac{1}{k}$; and the $l$ D8-brane intersection gives a novel
construction of an edge state between two abelian FQHEs with
different filling factions $\nu=\frac{1}{k}$ and $\frac{1}{k-l}$. We
then compute the Hall conductivity holographically. The D4-brane
computation correctly reproduces the standard FQHE result, while the
D8-brane result provides a prediction for the yet-to-be measured
Hall conductivity in such interfaces of quantum Hall fluids.

One may worry that realistic FQHE systems do not have any
 supersymmetry as opposed to the above examples. However, in the effective
 field theory description, the FQHE essentially occurs due to the
 presence of the Chern-Simons gauge field and its superpartners such as scalar fields
 and fermions do not contribute in any important way. Therefore, we can still
  capture the standard FQHE even in supersymmetric theories.

\paragraph{Model II: holographic realization of level-rank duality and
topological entanglement entropy.} FQHE is described by pure
Chern-Simons theories, whereas model I (which is based on
$\mathcal{N}=6$ Chern-Simons theory) has additional matter
fields.\footnote{This does not affect the
computation of the Hall conductivity since it is a topological
quantity.} Motivated by this, in the second model, we realize
holographically the pure Chern-Simons gauge theory.

Model II is based on the AdS/CFT for a D3-D7 system.\footnote{ For a
realization of the integer quantum Hall effect using a different
configuration of D3-D7 system, refer to the recent interesting work
\cite{DKS}.} We start with the familiar non-supersymmetric AdS/CFT
correspondence generated by $N$ D3-branes compactified on a circle
with anti-periodic boundary conditions for fermions: at low
energies, the boundary theory is a 3D bosonic Yang-Mills theory with
$U(N)$ gauge group; while the bulk side is the $AdS_5$ soliton of
type IIB string \cite{Witten}. We then deform this system by adding
$k$ D7-brane at the tip of the soliton. The boundary theory now
becomes a $U(N)_k$ Yang-Mills-Chern-Simons theory; and the bulk
geometry becomes a $AdS_5$ soliton with additional $k$ axion fluxes.
Finally we take the IR limit: at the boundary the Chern-Simons term
dominates over the Yang-Mills term; while in the bulk the theory
reduces to probe D7-brane worldvolume action. Therefore in the end
we arrive at a duality between the $U(N)_k$ Chern-Simons theory
living on $N$ D3-branes compactified on a circle with the $U(k)_N$
Chern-Simons theory living on the $k$ probe D7-brane compactified on
$S^5$. Interestingly, the holography in this D3-D7 system manifests
itself as the level-rank duality of the pure $U(N)_k$ Chern-Simons
theory by taking low energy limit.

In this model, we can holographically compute the Hall conductivity both from the bulk
supergravity and from the theory on a probe D3-brane or D7-brane which is dual to an
edge state. The topological
entanglement entropy is a quantity that can be computed from a given
ground state of the FQHE or the Chern-Simons effective field theory
\cite{KP,LW,DFLN}, and encodes information on the type of
quasi-particles (topological excitations) that can be build on the
ground state. We will clarify how the topological entanglement entropy
appears via the AdS/CFT duality \cite{Maldacena}.

\paragraph{Model III: holographic realization of hierarchical FQHE via
resolved $\mathbb{C}^2/Z_{n}$ singularity.} As mentioned earlier,
the single $U(1)$ Chern-Simons theory can only describe FQHE whose
filling fraction $\nu$ satisfies $\frac{1}{\nu}\in\mathbb{Z}$. To
model more generic (hierarchical) FQHEs we need to employ systems
with multiple $U(1)$ gauge fields. Inspired by the resemblance
between the continued fraction form of $\nu$ in hierarchical FQHEs
and the Hirzebruch-Jung continued fraction in the minimal resolution
of the (in general non-supersymmetric) orbifold $\mathbb{C}^2/Z_n$,
we consider the holographic duality between the F1-NS5 system on
$\mathbb{C}^2/Z_{n}$ and type IIA string in the $AdS_3\times S^3
\times \mathbb{C}^2/Z_{n}$ background. The FQH system now lives in
the bulk $AdS_3$ and the RR 3-form compactified on the chain of
blown-up 2-cycles (exceptional divisors) gives rise to a chain of
$U(1)$ gauge fields, corresponding to the hierarchy of $U(1)$ fields
in the effective description of the hierarchical FQHE.

Different from the previous two models, the FQH system in Model III
lives in the (2+1)-dimensional bulk geometry and its edge states are
at the boundary. The holographic duality in string theory is then
exactly the edge/bulk correspondence in the condense matter physics
context.

\paragraph{}
This paper is organized as follows. Model I, II and
III are presented in Section 2, 3 and 4, respectively. Section 5
summarizes the main merits of our constructions and discusses
several general issues in modeling FQHE using string theory in view
of recent developments of FQHE.

\section{Edge States in $\mathcal{N}=6$ Chern-Simons Theory}
\setcounter{equation}{0}

The purpose of this section is to define and holographically study
the supersymmetric version of edge states, which are particular
types of interfaces or defects, in the $\mathcal{N}=6$ Chern-Simons
theory. The $\mathcal{N}=6$ Chern-Simons theory consists of
bifundamental scalar fields and fermions in addition to the gauge
fields. Therefore it includes dynamical degrees of freedom. However,
the QHE is a rather topological phenomenon and these extra degrees
of freedom do not play important roles, as we will confirm from the
holographic calculations.

\subsection{Holographic Dual of $\mathcal{N}=6$ Chern-Simons Theory}

We consider the $AdS_4/CFT_3$ duality of the $\mathcal{N}=6$
Chern-Simons theory (ABJM theory) \cite{ABJM}. The duality can be
generated by $N$ M2-branes probing the singularity
$\mathbb{C}^4/Z_k$: the CFT side is the $\mathcal{N}=6$
superconformal
 Chern-Simons-Matter theory; the gravity side is the
M-theory living in $AdS_4\times S^7/Z_{k}$.

The $\mathcal{N}=6$ Chern-Simons theory consists of two gauge fields
$A^{(1)}$ and $A^{(2)}$ for the $U(N)_{k}\times U(N)_{-k}$ gauge
group in addition to bifundamental matter fields (scalars and
fermions). If we restrict to the gauge field sector, the action is
\begin{equation}
S^{gauge}_{\mathcal{N}=6}=\f{k}{4\pi}\int \mbox{Tr}A^{(1)}\we
dA^{(1)} -\f{k}{4\pi}\int \mbox{Tr}A^{(2)}\we dA^{(2)}.
\end{equation}
We represent this theory as $U(N)_k\times U(N)_{-k}$ Chern-Simons
theory, where $k$ and $-k$ denote the levels. In our application to
the quantum Hall effect, we will focus on the first $U(N)$ gauge
group.

Assuming $k$ is large, the gravity side is reduced to the type IIA string in $AdS_4\times
\mathbb{CP}^3$ \cite{ABJM}.
Before taking the $Z_k$ quotient, the gravity side is $AdS_4\times
S^7$:
\begin{equation}
ds^2_{11D}=\frac{R^2}{4}(ds^2_{AdS_4}+4ds^2_{S^7}),\qquad
\textrm{with} \qquad ds^2_{S^7}=(dy+A)^2+ds^2_{\mathbb{CP}^3}
\end{equation}
where $R=(2^5\pi^2kN)^{1/6}$ and $y\sim y+2\pi$. The metric of
$\mathbb{CP}^3$ is explicitly expressed as follows in the
coordinate system \cite{NTP}
\ba
ds^2_{\mathbb{CP}^3}&=&d\xi^2+\cos\xi^2\sin^2\xi\left(d\psi+\f{\cos\theta_1}{2}d\vp_1-
\f{\cos\theta_2}{2}d\vp_2\right)^2 \no &&
+\f{1}{4}\cos^2\xi\left(d\theta_1^2+\sin^2\theta_1
d\vp_1^2\right)+\f{1}{4}\sin^2\xi(d\theta_2^2+\sin^2\theta_2
d\vp_2^2). \label{cp} \ea

Now we take the $Z_k$ quotient: $\tilde{y}\equiv ky$ with
$\tilde{y}\sim \tilde{y}+2\pi$; and reduce to IIA via the reduction
formula (in this paper we always work with the string frame metric
setting $\al=1$)
\begin{equation}\label{11Dmetric}
ds^2_{11D}=e^{-2\phi/3}ds^2_{IIA}+e^{\f{4}{3}\phi}(d\ti{y}+\ti{A})^2,
\end{equation}
with $e^{2\phi}=\f{R^3}{k^3}= 2^{\f{5}{2}}\pi \s{\f{N}{k^5}}$ and
$\ti{A}=k A$. The RR 2-form $F^{(2)}=d\ti{A}$ in type IIA string (we
sometimes call this a D6-brane flux) is explicitly given as follows
\begin{eqnarray} F^{(2)}&=& k\Bigl(-\cos\xi\sin\xi d\xi \we
(2d\psi+\cos\theta_1d\vp_1-\cos\theta_2 d\vp_2)\no &&
-\f{1}{2}\cos^2\xi\sin\theta_1 d\theta_1\we d\vp_1
-\f{1}{2}\sin^2\xi\sin\theta_2 d\theta_2 \we d\vp_2\Bigr), \label{fluxsix}
\end{eqnarray}
while the RR 4-form remains the same:
$F^{(4)}=\f{3R^3}{8}\ep_{AdS_4}$, produced by the background D2-branes.

 The IIA string frame metric gives
the $AdS_4\times \mathbb{CP}^3$ IIA background \cite{ABJM}:
\begin{equation}
ds_{IIA}^2=\ti{R}^2(ds_{AdS_4}^2+4ds_{\mathbb{CP}^3}^2),
\end{equation} where
$\ti{R}^2=\f{R^3}{4k}=\pi\s{\f{2N}{k}}$. This background preserves
$24$ supersymmetries after the near-horizon supersymmetry
enhancement, which match the three-dimensional ${\cal N}=6$
superconformal symmetry in the CFT side. Note that in this
coordinate system the volumes of $\mathbb{CP}^1$ and $\mathbb{CP}^3$
are:\footnote{ Also note that the volume of the unit $S^7$ is
$\mbox{Vol}(S^7)=\f{\pi^4}{3}$.} \be \mbox{Vol}(\mathbb{CP}^1) =4\pi
\ti{R}^2,\qquad \qquad \mbox{Vol}(\mathbb{CP}^3)=\f{32}{3}\pi^3
\ti{R}^6. \ee

At finite temperature, the $AdS_4$ is replaced with the $AdS_4$
black hole
\be
ds^2=-\left(r^2-\f{r_0^3}{r}\right)dt^2+\f{dr^2}{r^2-\f{r_0^3}{r}}
+r^2(dx^2+dy^2). \label{xx}\ee The temperature is given by
$T=\f{3r_0}{4\pi}$.

\subsection{Edge States from Intersecting D-branes}

Now we consider the generalization of edge states in the QHE to
those in the pure $U(N)_k$ Chern-Simons theory. The edge state is
naturally defined by a line defect beyond which the rank $N$ of the
gauge group or its level $k$ changes. If we set $N=1$, then the
first type of the defect reduces to the ordinary edge state of the
QHE at least formally. The latter reduces to an edge which separates
the quantum Hall liquid with the different filling fractions
$\nu=\f{1}{k}$ and $\nu=\f{1}{k'}$. Recently this kind of interface
has been realized experimentally in \cite{expfilling} to detect the
fractional statistics in the FQHE.

In our $\mathcal{N}=6$ Chern-Simons theory with the gauge group
$U(N)_k\times U(N)_{-k}$, we can define the two types of
supersymmetric edge states in the same way by concentrating on the
first $U(N)$ gauge group; one shifts the rank, and the other shifts
the level. Indeed, we can find two D-brane constructions of edge
states in type IIA string on $AdS_4\times \mathbb{CP}^3$.

\begin{figure}[htbp]
   \begin{center}
     \includegraphics[height=5cm]{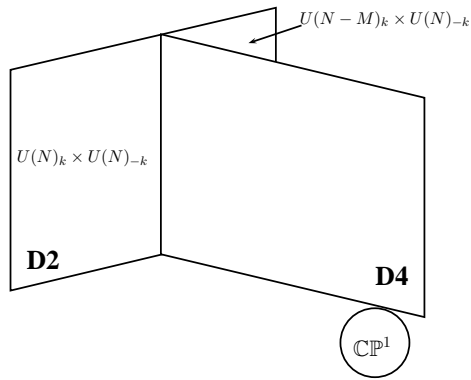}
   \end{center}
  \caption[Edge state from D4 intersecting D2]{Edge state from
intersecting $M$ D4-branes wrapped on $\mathbb{CP}^1$ with $N$
D2-branes: the gauge group on D2-branes changes from $U(N)_{k}\times
U(N)_{-k}$ to $U(N-M)_{k}\times U(N)_{-k}$ when crossing the edge.}
    \label{fig:D2D4}
\end{figure}

\begin{figure}[htbp]
   \begin{center}
     \includegraphics[height=5cm]{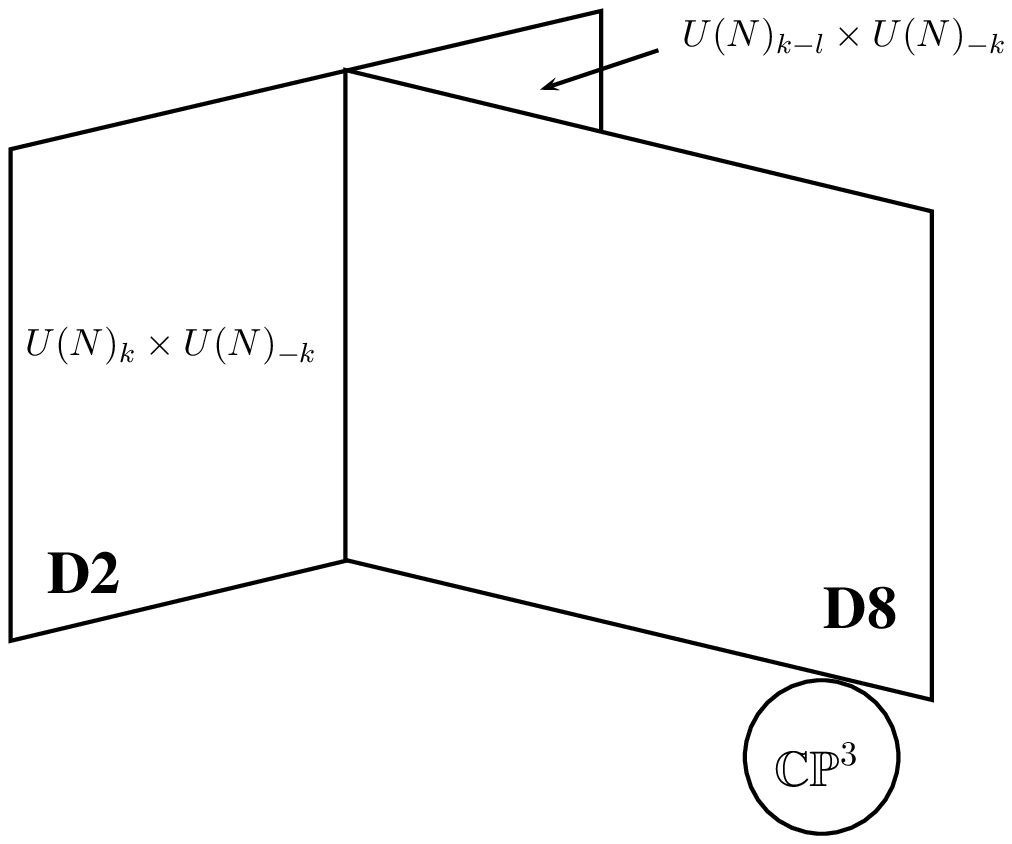}
   \end{center}
\caption[Edge state from D8 intersecting D2]{Edge state from
intersecting $l$ D8-branes wrapped on $\mathbb{CP}^3$ with $N$
D2-branes: the gauge group on D2-branes changes from $U(N)_{k}\times
U(N)_{-k}$ to $U(N)_{k-l}\times U(N)_{-k}$ when crossing the edge.}
\label{fig:D2D8}
\end{figure}

First, we can wrap a D4-brane on the $\mathbb{CP}^1$ inside
$\mathbb{CP}^3$ and attach it to a stack of $N$ D2-branes along one
spatial dimension, as shown in Fig.\ \ref{fig:D2D4}. Alternatively,
we can wrap a D8-brane on the entire $\mathbb{CP}^3$ and then
attach it to the D2-branes, as shown in Fig.\ \ref{fig:D2D8}. These
two kinds of intersections of the D4 and D8 with the D2-branes are
(1+1)-dimensional defects. We regard them as the string theory
descriptions of the two kinds of edge states in the $\mathcal{N}=6$
Chern-Simons theory. As we will see below, these two edge states
affect the boundary theory in different ways. If we consider an edge
state which consists of $M$ D4-branes, the rank of the gauge group
jumps from $U(N)_{k}\times U(N)_{-k}$ to $U(N-M)_{k}\times
U(N)_{-k}$. On the other hand, at the edge states defined by $l$
D8-branes, the level jumps from $U(N)_{k}\times U(N)_{-k}$ to
$U(N)_{k-l}\times U(N)_{-k}$. Both constructions break the parity
symmetry of the original ABJM theory.

The former argument is consistent with the interpretation of
fractional branes in \cite{ABJ}. If we consider an unstable
configuration of $M$ D4-brane wrapped on $\mathbb{CP}^1$ which are
parallel with the background D2-branes, then the D4-branes fall into
the horizon of $AdS_4$ and only the flux remains. This background is
argued to be dual to the $\mathcal{N}=6$ Chern-Simons theory with
the gauge group $U(N-M)_{k}\times U(N)_{-k}$ \cite{ABJ}.

The latter argument leads us to a new setup of the $AdS_4/CFT_3$ as
a byproduct. The IIA string on $AdS_4\times \mathbb{CP}^3$ with the
RR flux due to $l$ D8-branes wrapped on $\mathbb{CP}^3$ is dual to
the Chern-Simons theory with the gauge group $U(N)_{k-l}\times
U(N)_{-k}$, which preserves $\mathcal{N}=3$ supersymmetry.

Though we will not discuss below, we can also find other types of
similar brane configurations i.e. a D2-brane or D6-brane wrapped on
$\mathbb{CP}^3$. They are edge states which shift the rank or level
of both of the two $U(N)$ gauge fields simultaneously.

\subsubsection{D2$\bot$D4 Intersection}

We first consider the domain wall produced by a probe D4-brane that
wraps on the $\mathbb{CP}^1$ inside $\mathbb{CP}^3$ and attaches to
$N$ D2-brane along one spatial direction:
\begin{equation}\label{D2-D4}
\begin{array}{r|cccc|cccccccl}
\,\, \mbox{$AdS_4\times \mathbb{CP}^3$:}\,\,\, & t     & x
& y & r  & \theta_1 & \vp_1   & \theta_2   & \vp_2   & \xi   & \psi & \, \nonumber\\
\hline
N
\,\, \mbox{D2:}\,\,\,& \x &   \x  &  \x &   &  &   &  &  &
&
&\,\nonumber\\
k\,\, \mbox{D6-flux:}\,\,\,& \x &   \x  &  \x &   &  &   & \x & \x &
\x&\x
&\,\nonumber\\
1\,\, \mbox{D4:}\,\,\, & \x &  \x &   & \x  & \x & \x  &   & & & &\,
\end{array}
\end{equation}
where the coordinate $(\theta_1,\vp_1,\theta_2,\vp_2,\xi,\psi)$
labels the $\mathbb{CP}^3$. We choose the $\mathbb{CP}^1$ to be the
one defined by $(\theta_1,\vp_1)$ at $\xi=0$ so that the D4-brane in
$AdS_4\times \mathbb{CP}^3$ preserves 12 out of the total 24
supersymmetries. One can show this by lifting the system to M-theory
and counting the number of supersymmetries. The detailed proof is in
Appendix \ref{D4susy}.

This D4-brane has the world-volume action
\begin{equation}
S_{D4}=-T_{4}\int d^5\sigma e^{-\phi}\s{-\det(G+2\pi
F)}+2\pi^2T_4\int C^{(1)}\we F\we F,
\end{equation} where
$T_4=(2\pi)^{-4}$ in the unit $\al=1$ and $C^{(1)}$ is sourced by
the $k$ D6-flux (\ref{fluxsix}), which leads to
$\int_{\mathbb{CP}^1} F^{(2)}=2\pi k$. Integrating over the internal
$\mathbb{CP}^1$, the D4 Chern-Simons term becomes
\begin{equation}
S^{CS}_{D4} =\f{k}{4\pi}\int A\we dA.
\end{equation}
Note that the D4-brane wrapped on $\mathbb{CP}^1$ is charged only
under one (i.e. the first) of the two gauge groups $U(N)_k\times U(N)_{-k}$.

The probe D4-brane ends on a codimension-1 subspace of the boundary
$R^{1,2}$ of the $AdS_4$, thus creating a domain wall on the
$D2$-brane. The gauge group changes from $U(N)_k\times U(N)_{-k}$ to
$U(N-1)_k\times U(N)_{-k}$. The disappeared $U(1)$ is carried by the
D4-brane that extends into the bulk.

In order to relate to an ordinary QH system with a single gauge
group, we can concentrate on the first gauge group $U(N)_k$ of our
Chern-Simons theory, treating the second one with $U(N)_{-k}$ as a
spectator. Even though the two gauge fields are interacting in our
theory, this fact is not important when we consider the Hall
conductivity since it is quantized and determined by the topological
structure of the theory.

To understand the charge and statistics of a quasi-particle in the
$\mathcal{N}=6$ Chern-Simons theory, consider a fundamental string
(F-string) which is charged under one of the two gauge groups. To
extract one of the gauge groups, we insert a D4-brane wrapped on
$\mathbb{CP}^1$ (called the fractional D2-brane \cite{ABJ}) as a
probe and consider a F-string which ends at a point $(x_0,y_0)$ on
the D4-branes. Such a string can be found by considering the one
between the edge D4-brane and a D4 wrapped on $\mathbb{CP}^2$
(called dibaryon \cite{ABJM,ABJ}). This corresponds to a static
charge that generates $j^0=\delta(x-x_0)\delta(y-y_0)$ on the edge
D4-brane. The relevant part of the edge D4-brane action including
the F-string charge becomes
\begin{equation}
\f{k}{4\pi}\int A\we dA+\int j^0 A_0, \label{csaction}
\end{equation}
after integrated over  $\mathbb{CP}^1$. Solving the equation of
motion, we obtained the magnetic field and its flux:
\begin{equation}
F_{xy}=\f{2\pi}{k}\delta(x-x_0)\delta(y-y_0),\ \ \ \  \Phi=\int dxdy F_{xy}=\f{2\pi}{k}.
\end{equation}
Therefore, when two such quasi-particles are interchanged, the total
wave-function acquires a phase factor
$e^{i\frac{\pi}{k}}$.\footnote{Note that a factor of $\frac{1}{2}$
coming from the charge renormalization in the presence of the
Chern-Simons term cancels the factor of $2$ that follows from the
fact that a quasi-particle carries both charge and flux
\cite{Wen:1988cx}.} Namely, the F-string creates an anyon with the
fractional statistics. The same situation appears when we consider
the FQHE with the filling fraction
\begin{equation}
\nu=\frac{1}{k}.\label{filling}
\end{equation}
Indeed, the low energy effective theory of the FQHE with
(\ref{filling}) is known to be described by the Chern-Simons theory
at level $k$, namely (\ref{csaction}). The dibaryon (a D4 wrapped on
$\mathbb{CP}^2$) is interpreted as the quasi-particle (anyon). In
the ordinary FQHE, $k$ is taken to be an odd integer since an
electron obeys the fermionic statistics.

This means that a F-string should have the fractional charge
$\frac{e}{k}$ in the boundary quantum Hall system. Recall that in
string theory, the F-string has unit charge under gauge field. To
match with the experimental result, we rescale the gauge field:
$A\rightarrow \frac{e}{k}A$, under which the Chern-Simons terms
becomes
\begin{equation}
S^{CS}_{D4} =\frac{\ti{k}_{D4}}{4\pi }\int A\wedge dA,
\qquad \qquad \textrm{with} \qquad \ti{k}_{D4}\equiv \frac{e^2}{k}.
\end{equation}

To summarize, a D4-brane wrapped on $\mathbb{CP}^1$ and
attached to D2-brane creates an edge that divides the D2-brane into
two parts: the gauge group of its CFT is $U(N)_k\times U(N)_{-k}$ on
one side of the wall and $U(N-1)_{k}\times U(N)_{-k}$ on the other.
Though this is already an interesting edge state in a generalized
sense, to relate to the ordinary edge state in the QHE, one can
simply set $N=1$.\footnote{Usually in AdS/CFT we have to take $N$ to
be large enough  to make quantum corrections small. However, since
the QHE is a topological phenomenon (at zero temperature),
the essential results will not
be corrected by quantum fluctuations. Other quantities such as the
finite temperature correction to the Hall conductivity
 due to the thermal excitations of quasi-particles may receive substantial
 $1/N$ corrections. However, we expect no change to qualitative results.}
   Then treating the second
gauge field as a spectator, we can obtain a $U(1)_k$ Chern-Simons
gauge theory which models a quantum Hall system with filling
fraction $\nu=\f{1}{k}$; and the D2-D4 intersection describes the
edge state of this quantum Hall system.

Generalizing to multiple D4-brane systems, we find
that the edge between  $U(N)_k\times
U(N)_{-k}$ and $U(N-M)_{k}\times U(N)_{-k}$
 is dual to the $U(M)_k$
Yang-Mills-Chern-Simons theory.

Before we continue, let us compute the entropy of the D4-brane at
finite temperature. It is estimated as follows (we assume that it is
extended in $(t,x,r)$ direction with the interval $0<x<L$. We also
put the UV cut off at $r=r_{\infty}$) \be \beta
F=-S_{D4}=e^{-\phi}T_{4}\cdot L\cdot
\mbox{Vol}(\mathbb{CP}^1)\int^\beta_0 dt \int^{r_{\infty}}_{r_0} dr~
r. \ee This leads to the entropy \be S=\f{2}{9}\pi NLT. \ee We might
be able to compare with the standard result $S=\f{c}{3}\cdot \f{\pi
L}{\beta}$ in two dimensional CFT and find the effective central
charge $c=\f{2}{3}N$. These degrees of freedom represent the
massless modes of the open strings between $N$ $D2$ and the probe
D4-brane. It is interesting to note that this entropy is of the same
order as its free field theory result, in contrast to the bulk
entropy which deviates strongly from its free field result---by a
factor of $\s{\f{N}{k}}$ as noted in \cite{ABJM}.

\subsubsection{D2$\bot$D8 Intersection}

Now we consider the second domain-wall construction: a
D8-brane that wraps the entire $\mathbb{CP}^3$ and attaches to $N$
D2-brane along one spatial direction:
\begin{equation}\label{D2-D8}
\begin{array}{r|cccc|cccccccl}
\,\, \mbox{$AdS_4\times \mathbb{CP}^3$:}\,\,\, & t     & x   & y & r  &
\theta_1 & \vp_1   & \theta_2   & \vp_2   & \xi   & \psi & \, \nonumber\\
\hline N
\,\, \mbox{D2:}\,\,\,& \x &   \x  &  \x &   &  &   &  &  &
&
&\,\nonumber\\
k\,\, \mbox{D6-flux:}\,\,\,& \x &   \x  &  \x &   &  &   & \x & \x &
\x&\x &\,\nonumber\\1\,\, \mbox{D8:}\,\,\, & \x &  \x &   & \x  & \x
& \x  & \x  &\x & \x&\x &\,
\end{array}
\end{equation}
Same as the D4 case, this system preserves 12 supersymmetries.
To prove it, we can no longer use the strategy applied in D4 case
earlier, since there is no M-theory lift for D8-brane in massless
IIA theory. Instead, we can count the number of supersymmetries by
directly solving the Killing spinor equation. The detailed proof is
in Appendix \ref{D8susy}.

The D8-brane world-volume action is
\begin{equation}
S_{D8}=-T_{8}\int d^9\sigma e^{-\phi}\s{-\det(G+2\pi
F)}+2\pi^2T_8\int C^{(5)}\we F\we F,
\end{equation} where
$T_8=(2\pi)^{-8}$ in the unit $\al=1$ and $C^{(5)}$ gives the
background RR-flux $\int_{\mathbb{CP}^3} F^{(6)}=(2\pi)^5 N$
generated by $N$ D2 branes. Integrating over the internal
$\mathbb{CP}^3$, the D8 Chern-Simons term becomes
\begin{equation}
S^{c.s}_{D8} =\f{N}{4\pi}\int A\we dA.
\end{equation}

The analysis is similar to the D4 case. The intersection between the
probe D8-brane wrapped on $\mathbb{CP}^3$ and the stack of $N$
D2-branes creates a domain wall on the $D2$-brane. Since the
D8-brane background induces
 a Chern-Simons coupling of D2-branes,\footnote{This can be easily understood if
we recall that in massive IIA supergravity we have a RR 0-form field
strength sourced by the D8-branes. Notice that the 2-form RR flux
changes the difference of the levels between the two gauge groups,
keeping the sum unchanged. } the sum of levels of the first and
second gauge group jumps across the domain wall: i.e., the gauge
group $U(N)_k\times U(N)_{-k}$ changes into $U(N)_{k-1}\times
U(N)_{-k}$. Same as the D4 case, we will concentrate on the first
gauge group as the edge does not affect the second one.

Due to the same reason we explained for the D4 edge, the F-string
charge should be regarded as $\f{e}{k}$ in the FQHE interpretation.
By rescaling the gauge field $A\rightarrow \frac{e}{k}A$, we obtain
the D8 Chern-Simons term
\begin{equation}
S^{c.s}_{D8}=\frac{\ti{k}_{D8}}{4\pi }\int A\wedge dA,
\qquad \qquad \textrm{with} \qquad \ti{k}_{D8}\equiv \frac{N
e^2}{k^2}.
\end{equation}

If we set $N=1$ and ignore the $U(N)_{-k}$ part, we obtain an edge
state which sits between a $U(1)_k$ Chern-Simons gauge theory and a
$U(1)_{k-1}$ theory. This gives the edge state between two quantum
Hall liquids (QHL) with different filling factions:
$\nu=\frac{1}{k}$ and $\nu'=\frac{1}{k-1}$. Generalizing to multiple
D8-brane systems, we see that the edge between two QHL with
different filling fractions $\f{1}{k}$ and $\f{1}{k'}$ is dual to
the level-$1$ $U(k-k')$ Yang-Mills-Chern-Simons theory. More
generally, if we add both $M$ D4-branes and $l$ D8-branes to the $N$
D2-branes with $k$ D6-flux system, the theory changes into the
$U(N-M)_{k-l}\times U(N)_{-k}$ Chern-Simons theory when crossing the
edge.

Finally, by computing the thermal entropy of the CFT
\begin{equation}
S=\f{2\pi}{27}\f{N^2}{k}LT,
\end{equation}
we can find the effective central charge $c=\f{2N^2}{9k}$.

\subsection{Holographic Action of Edge State}

We analyze the holographic description of the edge states by studying the
D4 and D8-brane action. We assume the finite temperature background which is dual to
the $AdS_4$ black hole. We work with the following metric by setting $z=r^{-1}$ in (\ref{xx})
\begin{equation}\label{AdS4Poincare}
ds^2=-\f{H(z)}{z^2}dt^2+\f{dz^2}{H(z)z^2}+\f{dx^2+dy^2}{z^2},\ \ \ \
H(z)=1-r_0^3z^3.
\end{equation}
After integrating the internal $\mathbb{CP}^1$, the D4-brane
worldvolume action is:\footnote{We choose the convention that $\int
A\we dA=\int dtdxdz \ep^{ijk}A_i\de_j A_k$.}
\begin{equation}\label{dfaction}
S_{D4}=-\ap_{D4} \int d^3\sigma \s{-\det{(g_{ij}+2\pi F_{ij})}}
+\f{\ti{k}_{D4}}{4\pi}\int A\we dA.
\end{equation}
Similarly, a D8-brane wrapped on $\mathbb{CP}^3$
has the same action as (\ref{dfaction}), except that
$(\alpha_{D4},\tilde{k}_{D4})$ is replaced by
$(\alpha_{D8},\tilde{k}_{D8})$.

The values of coefficient $\alpha$ for D4-brane and D8-brane are
\begin{equation}
\ap_{D4}=\mbox{Vol}(CP^1)T_{4}e^{-\phi}\ti{R}^5=\f{N}{4\pi},\qquad
\ap_{D8}=\mbox{Vol}(CP^3)T_{8}e^{-\phi}\ti{R}^9=\f{N^2}{12\pi k}.
\end{equation}
The levels $\ti{k}$ of the Chern-Simons terms are
\begin{equation}
\ti{k}_{D4}=\frac{e^2}{k},\qquad \qquad
\ti{k}_{D8}=\frac{Ne^2}{k^2}.
\end{equation}

Since the D2$\bot$D4 and D2$\bot$D8
systems
provide the same action
to model the edge state, we will analyze the two systems
simultaneously using the action (\ref{dfaction}) with generic
parameter $(\alpha, k)$. Plugging their values for D4 or D8-brane
then produces the result for the corresponding system.

\subsubsection{Gauge Choice and Boundary Term}

There are two codimension-1 boundaries in our system: $z=0$ (the
position of the edge state) and $z=z_{h}=1/r_{0}$ (the position of
the horizon).\footnote{$z_{h}\rightarrow \infty$ at zero
temperature.} The $z=0$ boundary is particularly important;
\footnote{We will adopt the standard assumption in the holographic
computation of conductance: $\delta A|_{z=z_{h}}=0$
\cite{KB,KMMMT}.} as we will explain, it requires the addition of a
boundary term for the Chern-Simons action. The contribution from the
surface term is crucial to the holographic computation of the
conductivity.

The existence of boundary creates two problems. First, it breaks the
gauge invariance of the Chern-Simons term: under the gauge
transformation $\delta A= d\chi$
\begin{equation}
\delta_{\chi}S_{CS}=\frac{\tilde{k}}{4\pi}\int d\chi\wedge
dA=-\frac{\tilde{k}}{4\pi}\int_{\Sigma(z=0)} dtdx \chi F_{tx} \neq
0.
\end{equation}
This means that the gauge symmetry is preserved only if we restrict
to the subspace with the flat connection
$F_{tx}=0$ at $z=0$.

Second, the on-shell variation of Chern-Simons action has both
$\delta A_t$ and $\delta A_x$ term on the $z=0$ boundary:
\begin{equation}
\delta S|_{on-shell}=\frac{\tilde{k}}{4\pi}\int_{\Sigma (z=0)}
dtdx(-A_x \delta A_t+A_t \delta A_x)
\end{equation}
However, only one of the two terms should appear if $A_t$ and
$A_x$ are considered as a pair of canonical variables with
respect to $z$.

These two problems are improved simultaneously by adding a boundary
term as in \cite{EMSS} at $z=0$:
\begin{equation}\label{bdyaction}
S_{bdy}=\eta\frac{\tilde{k}}{4\pi}\int_{\Sigma (z=0)} dtdxA_t A_x
\qquad \qquad \textrm{with}\qquad \eta=\pm 1.
\end{equation}
First, after including the boundary term, the full action transforms
under $\delta A=d\chi$ as
\begin{equation}
\delta_{\chi} (S_{CS}+S_{bdy})=-\f{\ti{k}}{4\pi}\int_{\Sigma(z=0)}
dt dx \chi\left[(1+\eta)\partial_{t}A_x + (-1+\eta)\partial_xA_t
\right]. \label{gity}
\end{equation}
Even though we need to assume
$\partial_{t}A_x=0$ when $\eta=1$ (or $\partial_{x}A_t=0$ when $\eta=-1$),
we can still realize any non-vanishing electric flux
$F_{tx}\neq 0$.

The second problem is resolved similarly. The on-shell variation of
the total action contains $\delta A_t$ or $\delta A_x$ term:
\begin{displaymath}
\delta (S|_{on-shell}+S_{bdy})=\left\{
         \begin{array}{ll}
\displaystyle
           \frac{\ti{k}}{2\pi}\int_{\Sigma (z=0)} dtdx(A_t \delta A_x), & \qquad\hbox{$\eta=1$;} \\
\displaystyle
           -\frac{\ti{k}}{2\pi}\int_{\Sigma (z=0)} dtdx(A_x
\delta A_t), & \qquad \hbox{$\eta=-1$.}
         \end{array}
       \right.
\end{displaymath}
In our later computation of the Hall conductivity, we will set
 $\eta=-1$.

\subsection{Holographic Computation of Hall Conductivity via Perpendicular Edges}

\subsubsection{Set-up}
The left part of Fig.\ \ref{fig:Bending brane} shows the standard
experimental setup to measure the quantum Hall conductance: an
$\vec{E}$-field runs across the $y$-direction and the Hall current
flows along the two edge states in the perpendicular $x$-direction.
The right part depicts our D-brane configuration to model the two
edge states: the probe D4-brane or D8-brane bends into an
arch-bridge shape; its two ends are attached to the boundary CFT and
form the two edge states. Near the boundary $z=0$, the system looks
like a pair of parallel D4 and $\overline{\textrm{D4}}$ (or D8 and
$\overline{\textrm{D8}}$). In high temperature regime, the single
bending-brane will break into a pair of parallel D4 and
$\overline{\textrm{D4}}$ (or D8 and $\overline{\textrm{D8}}$) that
end at the horizon as in \cite{KMMMT}, but the analysis can be done
essentially in the same way.

\begin{figure}[htbp]
   \begin{center}
     \includegraphics[height=5cm]{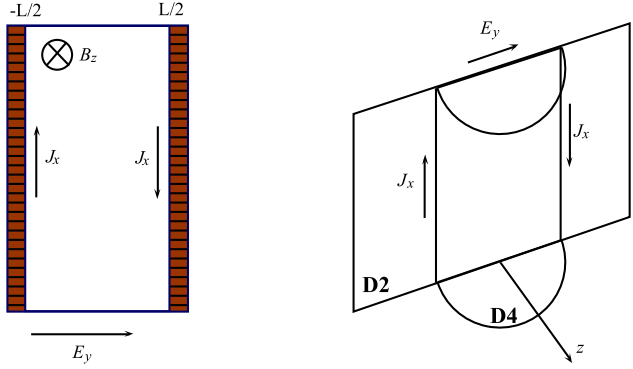}
   \end{center}
\caption[A bending brane that models the pair of edge states
parallel to $\vec{E}$-field]{ A bending brane that models the pair
of edge states perpendicular to $\vec{E}$-field.} \label{fig:Bending
brane}
\end{figure}

The probe bending D-brane is described by the world-volume
coordinates $(\sigma^0,\sigma^1,\sigma^2)=(t,x,y)$. Its profile is
specified by the function $z=z(y)$, with $-\f{L}{2}\leq
y\leq\f{L}{2}$. This is a holographic dual of the two parallel edge
states separated by the distance $L$.

Since we are interested in computing the Hall conductance, we do not
apply $\vec{E}$-field along the edge state (i.e. $E_x=0$). Thus we
can always choose
\begin{equation}
A_t=A_t(y), \qquad\qquad A_x=A_x(y), \qquad\qquad A_y=0.
\end{equation}
up to a gauge transformation. The probe D-brane action with the boundary term
(\ref{bdyaction}) is then
simply
\begin{eqnarray}
S&=&-\ap \int dt dx dy
\f{1}{z^3}\s{H+z'^2+z^4HF_{xy}^2-z^4F_{ty}^2}+\f{\ti{k}}{4\pi}\int
dtdxdy \left(A_x
\partial_y A_t-A_t
\partial_y A_x \right)\no
&& +\eta \f{\ti{k}}{4\pi}\left(\int_{\textrm{bdy-1}} dtdx
A_tA_x+\int_{\textrm{bdy-2}} dtdx A_tA_x\right)
\end{eqnarray}
with ``bdy-1" at $y=-\frac{L}{2}$ and ``bdy-2" at $y=\frac{L}{2}$.

Its equations of motion are
\begin{eqnarray}
&& \de_x\left(\f{\ap zHF_{xy}}{\s{D}}\right)-\de_t\left(\f{\ap
zF_{ty}}{\s{D}}\right)=0,\qquad \de_y\left(\f{\ap
zHF_{xy}}{\s{D}}\right)+\f{\ti{k}}{2\pi}F_{ty}=0,\no
&&\de_y\left(\f{\ap
zF_{ty}}{\s{D}}\right)+\f{\ti{k}}{2\pi}F_{xy}=0,\no &&
3z^{-4}\s{D}-\f{1}{\s{D}}(2HF_{xy}^2-2F_{ty}^2)+\de_y\left(\f{z'}{z^3\s{D}}\right)=0.
\label{eomymm}
\end{eqnarray}
where $D=H+z'^2+z^4HF_{xy}^2-z^4F_{ty}^2$. At zero temperature, it
allows analytical solutions as shown in the appendix \ref{BendingBraneExactSol}. However, to obtain the
Hall conductance, it is enough to compute the conserved charges. The
bulk conserved charges corresponding to the
translation symmetries of
 $A_t$ and $A_x$ are
\begin{equation}
Q_{A_t}=\alpha\frac{zF_{yt}}{\sqrt{D}}+\frac{\tilde{k}}{2\pi}A_{x},\qquad\qquad
Q_{A_x}=-\alpha\frac{zHF_{yx}}{\sqrt{D}}-\frac{\tilde{k}}{2\pi}A_{t}.
\end{equation}

\subsubsection{Fractional Quantum Hall Conductivity}
The conserved currents on the boundary are charge density $\rho$ and
current density $j$. Following the arguments \cite{KB,KMMMT}, they are defined as
\begin{equation}
\rho=\f{\delta S}{\delta A_t}\Bigl|_{z=0}, \qquad \qquad j=\f{\delta
S}{\delta A_x}\Bigl|_{z=0}. \label{boundaryj}
\end{equation}
In the holographic computation of the currents, we need to specify
the boundary interaction. Recall that the two choices of $\eta$
correspond to different boundary conditions of $A$:
\begin{eqnarray}
&&\delta A_x|_{bdy}=0 \qquad \textrm{and} \qquad \delta A_t |_{bdy}
\,\,\,\textrm{free} \qquad \qquad \textrm{if}\quad \eta=1\no
&&\delta A_t |_{bdy}=0 \qquad \textrm{and} \qquad \delta A_x |_{bdy}
\,\,\, \textrm{free} \qquad \qquad \textrm{if}\quad \eta=-1.
\nonumber
\end{eqnarray}

Since we are interested in computing the Hall conductance, we need
to choose
\begin{equation}
\eta=-1
\end{equation}
so that on the boundary, $A_x$ is allowed to vary and $A_t$ is
fixed. The boundary values of $A_t$ correspond to the electric static
potential and their difference is determined by the $\vec{E}$-field:
\begin{equation}
A_t|_{bdy-2}-A_t|_{bdy-1}= V_y=E_yL
\end{equation}
There is an additional subtlety for high temperature regime: as the
temperature is increased, the horizon $z=z_h$ moves closer to the
tip of the bending-brane and eventually forces the single
bending-brane to break into a pair of parallel D-branes that end at
the horizon. In such a case, we need to also specify the boundary
condition for the $z=z_h$ boundary of the probe D-brane. As claimed
in \cite{KMMMT,KB}, the most natural one is simply $\delta
A|_{z=z_h}=0$: the boundary condition at the horizon does not
influence the physics at the spatial infinity.
We will adopt this argument throughout this work.

The charge and current density at each edge are then
\begin{equation}
\rho=Q_{A_t},\qquad \qquad
j_x=Q_{A_x}+\frac{\tilde{k}}{2\pi}A_t|_{bdy} \label{hcur}.
\end{equation}
Since the edge is (1+1)-dimensional, the current is equal to the
current density: $I=j$. What is measured experimentally is the net
current along both edges; and since currents from the two edges flow
in opposite directions, the net current along $x$-direction is
\begin{eqnarray}
I_{x}=I_{bdy-2}-I_{bdy-1}=j_{bdy-2}-j_{bdy-1}=\frac{\tilde{k}}{2\pi}
(A_t|_{\textrm{bdy-2}}-A_t|_{\textrm{bdy-1}})=\frac{\tilde{k}}{2\pi}V_y
\end{eqnarray}
This gives the Hall conductance
\begin{equation}\label{conductance}
G_{xy}=\frac{I_x}{V_y}=\frac{\tilde{k}}{2\pi}=\frac{\ti{k}}{h}.
\end{equation}
In the last step we restored $\hbar$ which has been set to $1$.

Note that (\ref{conductance}) is the conductance for the entire
(2+1)-dimensional quantum Hall system, and in this dimension, the
conductance is equal to the conductivity:
\begin{equation}\label{conductivity}
\sigma_{xy}=\f{j_x}{E_y}=\f{j_x\cdot L}{E_y\cdot L}=\f{I_x}{
V_y}=G_{xy}=\frac{\tilde{k}}{h}.
\end{equation}

Now we translate the result (\ref{conductivity}) into the quantum
Hall effect. For D4-brane case, recall that the D4-brane models the
edge states of a QHL with filling fraction $\nu=\frac{1}{k}$ and
plugging $(\ti{k},\nu)_{D4}=(\frac{e^2}{k},\frac{1}{k})$ into
(\ref{conductivity}) reproduces the correct fractional quantum Hall
conductivity:\footnote{One may compare this with the quantization of
the conductance in quantum wire: $G\in \f{e^2}{h}{\mathbb{Z}}$.}
\begin{equation}
\sigma_{xy}=\f{\nu e^2}{h}.
\end{equation}
Notice that our holographic calculation of the quantum Hall
conductivity does not depend on the temperature. This is expected
since in the Chern-Simons description of the QHE the excitation gap
is infinitely large thus no quasi-particle (including excitation
into higher Landau levels) can  be thermally excited. This is the
reason that
we can measure the quantized Hall conductivity
with remarkable accuracy.

On the other hand, the $l$ D8-branes model the edge
state that separates two QHL with different filling fractions
$\nu=\frac{1}{k}$ and $\nu'=\frac{1}{k-l}$, setting $N=1$.
By preparing two edges as in the D4 case, we can realize a setup
in which the QHL with $\nu$ is surrounded by the QHL with $\nu'$
from the both left and right sides. Then we can consider
 the Hall conductivity $\ti{\sigma}_{xy}$ for the QHL with $\nu$
under this circumstance. Plugging
$\ti{k}_{D8}=\frac{e^2}{k^2}$ into (\ref{conductivity}) then
predicts that
\begin{equation}
\ti{\sigma}_{xy}=\f{le^2}{2\pi k^2}=\f{l\nu^2 e^2}{h}\simeq \f{|\nu'-\nu| e^2}{h},
\end{equation}
where we took into account that $k$ is taken to be large in the supergravity
description of the $AdS_4/CFT_3$ duality. This result can be naturally understood if
we remember the difference of the current between the left and right of the interface
does contribute to the conductivity in these generic examples.

It is also possible to see that the lon

\subsection{Alternative Computation of Hall Conductivity via Parallel Edges}

\subsubsection{Set-up: Edge D-brane Parallel to $\vec{E}$}

\begin{figure}[htbp]
   \begin{center}
     \includegraphics[height=5cm]{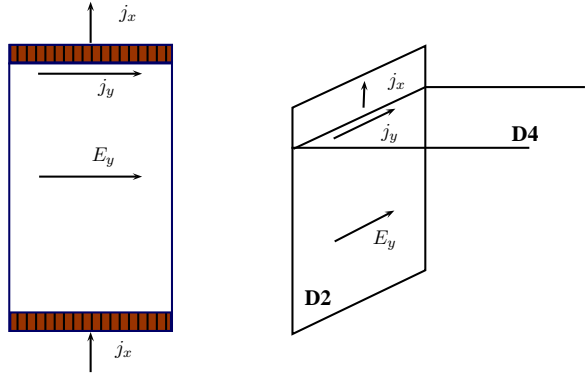}
   \end{center}
\caption[An infinite D4-brane that models the single edge that is
parallel to the $\vec{E}$-field.] {An infinite brane that models the
single edge that is parallel to the $\vec{E}$-field.}\label{fig:SingleEdge}
\end{figure}

In the previous subsection, we obtain the current density $j_{x}$ by
computing the current flowing \emph{along} the two edges
perpendicular to the field $E_y$. Experimentally, one actually
measures $j_x$ as the current density flowing \emph{out of} the edge
parallel to $E_y$, as shown in the left part of Fig.
\ref{fig:SingleEdge}. Therefore we can also construct an edge state
parallel to $E_y$, and compute $j_x$ as the current density
perpendicular to the edge state.\footnote{In this way, we can
calculate the Hall conductivity by looking at either
the horizontal edges or (equivalently)
the vertical edges. In the ordinary QHE context, we
can find these two different approaches implicitly in e.g.
\cite{Yoshioka}. This can be regarded as a kind of duality similar
to open-closed channel duality.}

Fig. \ref{fig:SingleEdge} illustrates the brane construction of the
edge state parallel to the $\vec{E}$-field. The graph on the left
side highlights the two edge states parallel to the $\vec{E}$-field.
The current $j_y$ flows along these edge states and $j_x$ flows
across them. The graph on the right depicts an infinite D4-brane
extending to the horizon and intersecting with D2-brane along
$y$-direction: this (1+1)-dimensional intersection models the edge
state shown on the left.

The continuity equation relates $j_x$ to $\rho$ (charge density
inside the edge) and $j_y$ (current density along the edge):
\begin{equation}
j_x=-\left(\f{\de\rho}{\de t}+\f{\de j_y}{\de y}\right). \label{cur}
\end{equation}
To compute $\rho$ and $j_y$ holographically, we can model this edge
by a single D4-brane (or D8-brane) wrapped on $\mathbb{CP}^1$, intersecting
with D2-brane along $y$-direction. This construction is actually
simpler than the previous one: since we only need to compute the
current flowing out of \emph{one} edge, it's enough to have a single
brane that extends all the way to the horizon. Notice that even
in this case we assume there are two edges which connect in the bulk or
which end at the black hole horizon so that the electric
current comes from one of them and
flows into the other.

The D4-brane world-volume action is
\begin{equation}\label{dfactionn}
S_{D4}=-\ap \int dt dy dz
\f{1}{z^3}\s{1+z^4H(z)F_{yz}^2-z^4F_{tz}^2-z^4H(z)^{-1}F_{ty}^2}
+\f{\ti{k}}{4\pi}\int A\we dA.
\end{equation}
The full equations of motion are
\begin{eqnarray} &&
\de_y\left(\f{\ap zHF_{yz}}{\s{\ti{D}}}\right)-\de_t\left(\f{\ap
zF_{tz}}{\s{\ti{D}}}\right)+\f{\ti{k}}{2\pi}F_{ty}=0,\no &&
\de_z\left(\f{\ap zHF_{yz}}{\s{\ti{D}}}\right)+\de_t\left(\f{\ap
zH^{-1}F_{ty}}{\s{\ti{D}}}\right)+\f{\ti{k}}{2\pi}F_{tz}=0,\\ &&
\de_z\left(\f{\ap zF_{tz}}{\s{\ti{D}}}\right)+\de_y\left(\f{\ap
zH^{-1}F_{ty}}{\s{\ti{D}}}\right)+\f{\ti{k}}{2\pi}F_{yz}=0.
\nonumber\label{eomym}
\end{eqnarray} where $\ti{D}\equiv
1+z^4H(z)F_{yz}^2-z^4F_{tz}^2-z^4H(z)^{-1}F_{ty}^2$. At zero
temperature, it allows analytical solutions as shown in Appendix
\ref{SingleBraneExactSol}.

\subsubsection{Hall Conductivity from A Single Edge}

The variation of the on-shell action gives the boundary currents
\be
\rho=-\f{\ap zF_{tz}}{\s{\ti{D}}}+\f{\ti{k}}{2\pi}A_y,\ \ \ \ \ \ \ \ \ \
j_y=\f{\ap zF_{yz}}{\s{\ti{D}}},
\ee
almost in the same way as we did in (\ref{hcur}).
 Note that for the $z=z_h$ boundary of the probe
D-brane, we again used the assumption $\delta A|_{z=z_h}=0$ as in
\cite{KMMMT}.

Then we plug these into (\ref{cur}) and finally we obtain by using (\ref{eomym})
 \be
j_x=-\f{\ti{k}}{2\pi}\de_y A_t=\f{\ti{k}}{2\pi}E_y, \ee where we
assumed $\de_t A_y=0$ using the argument of gauge invariance in
(\ref{gity}). This correctly reproduces the Hall conductivity \be
\sigma_{xy}=\f{\ti{k}}{2\pi}=\f{\ti{k}}{h}.\label{recon} \ee Thus we
reproduced the previous results (\ref{conductivity}) from this
independent argument.

\section{Holography of Pure Chern-Simons Theory and Topological Entanglement Entropy}

So far we have studied the supersymmetric Chern-Simons theory since
it can be holographically realized in a clear way. This theory
includes matter fields in addition to the Chern-Simons gauge theory
and is a dynamical CFT. However, if we have in mind the application
to topological insulators in condensed matter physics, we need a mass
gapped gauge theory. In this section we will consider a
(2+1)-dimensional gauge theory which flows into a pure Chern-Simons
gauge theory. We will also present its holographic dual. A
holographic background dual to a pure $N=1$ Chern-Simons model has
already been given in \cite{MN} and our model shares several similar
properties such as the level-rank duality.

\subsection{A Holographic Dual to Pure Chern-Simons and Level-Rank Duality}

We start with the $\mathcal{N}=4$ super Yang-Mills in four dimensions
and compactify one of its three spatial directions (denoted by
$\theta$). Then we impose the anti-periodic boundary condition on
$\theta$, which makes the fermions massive. The quantum corrections
give a mass to scalar fields and in the end we obtain a pure
Yang-Mills theory in $(2+1)$ dimensions in the IR limit. This theory
is dual to the $AdS_5$ soliton (or double Wick-rotated $AdS_5$ black
brane) in IIB string theory \cite{Witten} \be
ds^2=R^2\f{dr^2}{f(r)r^2}+\f{r^2}{R^2}
(-dt^2+f(r)d\theta^2+dx^2+dy^2)+R^2d\Omega_5^2, \label{hdcs} \ee
where $f(r)=1-\left(\f{r_0}{r}\right)^4$. The coordinate $\theta$ is
compactified as $\theta\sim \theta+L$, where $L=\f{\pi R^2}{r_0}$,
so that the total geometry becomes smooth. The cycle $\de_\theta$
shrinks at $r=r_0$. Thus the two-dimensional space spanned by
$(r,\theta)$ is topologically a two-dimensional disk $D_2$. The
boundary of the disk is at $r=\infty$ and this is the boundary of
the AdS. The dual gauge theory lives in $R^{1,2}$ whose coordinate is
$(t,x,y)$.

Now we would like to deform this theory so that it includes the
Chern-Simons term. Remember the WZ-term of a D3-brane, \be
\f{1}{4\pi}\int_{D3}\chi F\we F=-\f{1}{4\pi}\int_{D3}d\chi \we A\we
F, \ee where $\chi$ is the axion field in IIB theory. If we assume
the background axion field $\chi=\f{k}{L}\theta$, this leads to the
CS coupling \be \f{k}{4\pi}\int_{R^{1,2}} A\we F. \ee In this way,
we get a Maxwell-Chern-Simons theory which flows into the level $k$
pure Chern-Simons theory in the IR. This argument can be easily
generalized to $N$ D3-branes and we get the non-abelian Chern-Simons
term $\f{k}{4\pi}\int_{R^{1,2}} \mbox{Tr}\left(A\we dA+\f{2}{3}A^3\right)$.

\begin{figure}[htbp]
   \begin{center}
     \includegraphics[height=5cm]{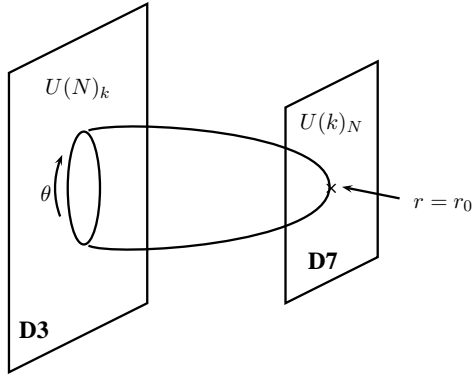}
   \end{center}
\caption[D3-D7 construction that generates a level-rank duality.]
{D3-D7 construction that generates a level-rank duality system: the
D3-brane wraps the $\theta$-direction and generates a $U(N)_k$
Yang-Mills-Chern-Simons theory. The bulk geometry is a $AdS_5$
soliton with additional $k$ axion flux, which are sourced by $k$
D7-branes sitting at $r=r_0$ where $\theta$-circle shrinks to zero.
The IR limit of D3-brane theory ($U(N)_k$ Chern-Simons theory) is
dual to the D7-brane theory which is a $U(k)_N$ Chern-Simons
theory.}\label{fig.D3D7}
\end{figure}

The axion field $\chi=\f{k}{L}\theta$ can be regarded as $k$
D7-branes located at $r=r_0$. Therefore we find that the $U(N)_k$
Yang-Mills-Chern-Simons theory is holographically dual to this
$AdS_5$ soliton background with the $k$ D7-branes (see  Fig.
\ref{fig.D3D7}). The IR limit of the CFT side is dual to the small
$r$ region of the dual background and this is given by the D7-brane
theory. Due to the Chern-Simons term $\f{N}{4\pi}\int_{R^3} A\we F$
from the RR 5-form flux, we get $U(k)_N$ pure Chern-Simons theory.
Thus this holography in the low energy limit is equivalent to the
level-rank duality of the Chern-Simons theory. Indeed, the
level-rank duality becomes more direct for $U(N)_k$ than for
$SU(N)_k$ as shown in \cite{NS}. Indeed, we can prove the following
identity for the partition functions on $S^3$ \cite{NS}: \be
Z(S^3,U(N)_k)=Z(S^3,U(k)_N),\ \ \ \  \ \
Z(S^3,SU(N)_k)=\s{\f{k}{N}}Z(S^3,SU(k)_N). \ee

The reason that the theory on the D7-branes should be regarded as a
pure Chern-Simons theory without the Yang-Mills term can be
understood by looking at the excitations on them. In the dual gauge
theory side (the D3-brane theory), the mass gap due to the
Kaluza-Klein compactification is given by \be
m_{KK}\sim\f{1}{L}\sim\f{r_0}{R^2}. \ee On the other hand, the
gravity theory is given by the IIB supergravity and the D7-brane
theory. In the presence of the ordinary Yang-Mills term obtained
from the D-brane action, the D7-brane theory has a mass gap due to
the Chern-Simons term, which is estimated as\footnote{ Here both
masses in D7 and D3-brane theory are measured by the same time $t$.}
\be m_{gauge}\sim N\f{r_0\al^2}{R^6}\sim \f{m_{KK}}{g_{YM}^2} \gg
m_{KK} \ee Thus the excitation on the D7-branes can be neglected and
it reduces to the pure Chern-Simons theory.

\subsection{Relation to Quantum Hall Effect}

Now we would like to examine the gauge theory dual to the gravity
background (\ref{hdcs}) with $k$ D7-branes.\footnote{It might be
useful to compare our model with the one by \cite{DKS}, which is
also constructed by putting D7-branes as probes in the $AdS_5\times
S^5$. In this model \cite{DKS}, the D7-branes extend to the infinity
and thus the dual QFT side includes massless fermions. Also the dual
$\mathcal{N}=4$ super Yang-Mills
 lives in $(3+1)$ dimension and the fermions live in its (2+1)-dimensional defect, where we expect the
 QHE. However, the holographic dual QFT of our model is purely (2+1)-dimensional (neglecting
 the Kaluza-Klein excitations). Also the D7-brane is localized in the IR region of AdS and thus it
corresponds to a normalizable deformation of the dual QFT. Therefore
no extra degrees of freedom is induced by the D7-branes. }

To interpret our model as one for the quantum Hall effect, we need
to couple it to an external gauge field. To this end, we assume that
the background RR 2-from field has the form \be B_{RR}= A_{RR}\we
d\ti{\theta}, \label{brr} \ee where $A_{RR}$ is a 1-form in
$R^{1,2}$ and will serve as the external gauge field. We define
$\ti{\theta}\equiv\f{2\pi}{L}\theta$ so that the period of
$\ti{\theta}$ is $2\pi$.

 The relevant RR-coupling on D3-branes reads
\be \f{1}{4\pi^2}\int_{D3} B_{RR}\we \mbox{Tr}F
=\f{1}{2\pi}\int_{R^{1,2}}A_{RR}\we \mbox{Tr}F. \ee Therefore in the
IR limit, the QFT side becomes the Chern-Simons theory coupled to an
external gauge field \be S_{D3}=\f{k}{4\pi}\int \mbox{Tr} \left(A\we
dA+\f{2}{3}A\we A\we A\right) +\f{1}{2\pi}\int  A_{RR}\we \mbox{Tr}
F, \label{HCS} \ee This is a non-abelian version of the
standard Chern-Simons description of the quantum Hall effect.

It is also important to identify the anyons in this system. A
D5-brane wrapped on $S^5$ is the baryon vertex in $AdS_5\times S^5$
\cite{WB}. Due to the string creation, it comes with $N$ F-strings
stretching between the D5 and $N$ D3-branes. The F-strings induce a
unit charge for each $U(1)$ gauge theory in
 $U(1)^N \subset U(N)$.
Thus according to the same argument as in
 (\ref{csaction}), when two baryons are interchanged, the full wave-function acquires
 a phase factor $e^{\f{\pi i N}{k}}$,
 which means that the baryons are anyons for generic values of $N$.

Now we would like to compute the Hall conductivity in the
holographic dual gravity side. Usually, the D7-branes extend to the
boundary of AdS \cite{Karch:2002sh}. However, in our case, they are
localized at the IR region $r=r_0$. Therefore we can replace them
with a background axion flux $\chi=\f{k}{L}\theta$ instead of
treating them as probe D-branes. Below we are only interested in the
leading order effects produced by this axion flux so we will not
take the backreaction to the metric, dilaton and fluxes into
account. Please refer to Appendix. \ref{HallD3D7} for the
calculation of Hall conductivity in the probe D7-brane analysis.

The holographic current is computed as the on-shell variation of the
action:\be j^\mu=\f{\delta S_{IIB}}{\delta A_{RR\mu}}\Bigr|_{bdy},
\label{condjk} \ee where $A_{RR}$ is the 1-form that comes from the
RR 2-form field $B_{RR}$ via (\ref{brr}) and serves as the external
gauge field in the D3-brane system. $S_{IIB}$ is the type IIB
supergravity action evaluated on-shell in the presence of D7-brane
flux.

After we integrate over $S^5$, the action which involves the $NS$
and $RR$ 3-form flux $H_{NS}$ and $H_{RR}$ can be written as follows
(in the convention of \cite{Pol} with $\al=1$)
 \be S_{Hflux}
=-\f{R^5}{2^4(2\pi)^4}\int d^5x\s{-G}
\left(g_s^{-2}|H_{NS}|^2+|H_{RR}-\chi H_{NS}|^2\right)
-\f{N}{(2\pi)^3}\int B_{NS}\we H_{RR}, \ee where the final term
comes from the Chern-Simons term $-\f{1}{4\kappa^2}\int C_4\we
H_{NS}\we H_{RR}$ together with the cross terms in
$-\f{1}{8\kappa^2}\int|\ti{F}_{5}|^2$ after compactifying on $S^5$
with 5-from flux (see also \cite{AW}).

Since the background field of quantum Hall effect on D3-branes is
$F_{RR}=dA_{RR}$, we assume $B_{NS}=0$ in the solution we are looking for.
This is consistent with equations of motion if
\be
*_5 H_{RR}=\f{32\pi^2N}{R^5 k}F_{RR},
\ee to the leading order of $k$; here $*_5$ denotes the Hodge dual
in the 5D spacetime. Plugging this into the remaining action of
$H_{RR}$: $S_{H_{RR}}=-\f{R^5}{2^4(2\pi)^4}\int H_{RR}\wedge *_5
H_{RR}$ and integrating over $\theta$, we finally obtain \be
S_{IIB}=\f{N}{4\pi k}\int F_{RR}\wedge F_{RR}. \ee Since this
topological term leads to the boundary Chern-Simons term in the
AdS/CFT procedure \be S_{bdy}=\f{N}{4\pi k}\int A_{RR}\wedge F_{RR},\label{CSef}
\ee the $x$-component of (\ref{condjk}) gives $j_x=\f{N}{ kh}E_y$.
Thus we find the fractional quantum Hall conductivity \be
\sigma_{xy}=\f{N}{kh}. \label{fincon} \ee This clearly agrees with
what we find from (\ref{HCS}). Note that the quantization of
D7-brane and D3-brane charges lead to the fractional quantization of
the Hall conductivity in (\ref{fincon}).

In this formalism, we can also easily see from (\ref{CSef})
 that the longitudinal conductivity vanishes
\be
\sigma_{xx}=0,
\ee
as expected in the standard FQHE.

\subsection{Topological Entanglement Entropy}

The topological entanglement entropy $S_{top}$ \cite{KP,LW} is
defined as the finite part of the entanglement entropy \be
S_{A}=\gamma\f{l}{a}+S_{top}\ , \label{tope} \ee where $a$ is the UV
cutoff and $\gamma$ a certain numerical factor proportional to the
number of UV degrees of freedom. The entanglement entropy $S_A$ is
defined as the von-Neumann entropy when we trace out a subsystem $A$
on a time-slice. We assume $A$ is a two-dimensional disk in a
time-slice. $l$ is the length of the boundary $\de A$ of $A$. When
the theory has a mass gap, the quantity $S_{top}$ can be shown to be
invariant under a continuous deformation of the region $A$
\cite{KP,LW}.

Conformal field theories such as the $\mathcal{N}=6$ Chern-Simons
theory do not have any mass gap, so the finite part of $S_A$ for
these theories is not a topological invariant. In contrast, the pure
Yang-Mills theory is a good example where we can define $S_{top}$
non-trivially. Indeed we can show that it becomes the log of the
partition function $Z(S^3)$ of Chern-Simons theory on $S^3$
\cite{EMSS} using the standard surgery method \cite{WiCS}.

On the other hand, the holographic calculation of topological entanglement
entropy has recently been attempted in \cite{PP} employing the holographic
formula of the entanglement entropy \cite{RT}.
 However, $S_{top}$ for the
(2+1)-dimensional Yang-Mills theory without
 Chern-Simons term turns out to be trivial. Therefore here we would like to see how the
holographic computation of
 the topological entanglement entropy is modified in the presence of the Chern-Simons term.

In the holographic calculation \cite{RT}, we introduce the
(negative) deficit angle $\delta=2\pi(1-n)$ on the $\de A$ at
$r=\infty$. Then we extend this deficit angle surface toward the
bulk AdS, called $\gamma$. Since the entanglement entropy is defined
by $S_{A}=-\f{\de}{\de n}\log Z_n$, where $Z_n$ is the partition
function on the manifold with the deficit angle $\delta=2\pi(1-n)$,
in our setup we obtain \be
S_A=\f{\mbox{Area}(\gamma)}{4G_N}+S_{D7}, \label{hole} \ee where the
part $S_{D7}$ is the contribution of $k$ D7-branes. The action
principle tells us that $\gamma$ is the minimal area surface
\cite{Fu,RT} whose boundary coincides with $\de A$.
It is clear from this holographic expression (\ref{hole}) that the topological entanglement entropy
is given by $S_{D7}$ as the Chern-Simons term appears due to the D7-branes.

Since we are interested in the (2+1)-dimensional
 Yang-Mills-Chern-Simons theory, the Kaluza-Klein modes need to be negligible.
This requires that the size of region $A$ be much larger than the
compactified radius $L$. When there is no D7-brane, the surface
$\gamma$ ends at the bubble wall $r=r_0$ as shown in \cite{NT,KKM,PP}.
In our case with D7-branes, and with their backreactions neglected,
the deficit angle surface $\gamma$ is extended into the
(2+1)-dimensional theory on the $k$ D7-branes. Since the large $N$
limit corresponds to the large level limit, the D7-brane theory
becomes the classical Chern-Simons theory. Then, the AdS/CFT tells
us that $S_{top}$ is given by $S_{D7}$, i.e. the topological
entanglement entropy  of $U(k)_N$ Chern-Simons theory. Using the
level-rank duality, this reproduces the $S_{top}$ of the original CS
theory. In this way, we find how to reproduce the correct
topological entanglement entropy via AdS/CFT.

If we reduce the size of the subsystem $A$, eventually, the surface
$\gamma$ separates from the the tip $r=r_0$. In the absence of
D7-branes, this is interpreted as the confinement/deconfinement
transition as the smaller size of $A$ probes more UV region of the
gauge theory \cite{NT,KKM}. Indeed the derivative of the
entanglement entropy $S_A$ jumps under this transition. In our case
with D7-branes, the topological term $S_{top}$ also disappears when
$A$ gets enough small. It is interesting to confirm by direct
calculations in specific models.

It would also be intriguing to find the backreacted geometry and
compute $S_{top}$ from purely bulk calculations. However, this seems
to be not straightforward since the direct Chern-Simons result show
the behavior (using the series expansion formula in \cite{Pe} of
$Z(S^3)$) \be S_{top}\sim -\f{k^2}{2}\log N, \ee in the large $N$
limit with $k$ kept finite. This suggests that this is not obtained
in the tree level supergravity. Finally we would like to mention
that we can discuss the holographic Wilson-loops
in exactly the same way.

\subsection{Edge States}

Since our holographic model is completely gapped in the low energy,
it is interesting to see how to add
some dynamical degrees of freedom in the UV. One of the best
examples are the edge states, which we have already discussed in the
$\mathcal{N}=6$ Chern-Simons theory. The edge state which appears in
the interface between $U(N)_k$ and $U(N+1)_k$ is described by a
D3-brane which stretches from $r=\infty$ to $r=r_0$ and wraps the
$\theta$ cycle. On the other hand, the edge which appears in the
interface between $U(N)_k$ and $U(N)_{k+1}$ is described by a
D7-brane which stretches from $r=\infty$ to $r=r_0$ and wraps the
$S^5$.

First consider the edge state generated by D3-branes. In the dual
gravity side, the $k$ D7-branes are regarded as the background axion
source; thus the holographic edge state is described by the $U(1)_k$
DBI-Chern-Simons theory living on the edge D3-brane that wraps on
$\theta$. The Hall conductivity can then be computed in the same way
as in section 2. In contrast to the pure Chern-Simons theory living
on the probe D7-branes that are localized at the tip $r=r_0$ and
wrapped on $S^5$, this D3-brane theory has propagating degrees of
freedom since the probe D3-brane extends to the boundary of the AdS.
Similarly, in the case of the edge state generated by D7-brane, due
to the presence of $F^{(5)}$ flux, the dual theory is a $U(1)_N$
DBI-Chern-Simons theory living on the D7-brane wrapped on $S^5$.

\section{Hierarchical Fractional QHE from String Theory}

So far we have only discussed the Chern-Simons gauge theory with a
single gauge group. However, it can only describe those fractional
QH systems with filling fractions in the form of
\begin{equation}
\nu=\f{1}{k},
\end{equation}
where $k$ is an integer and corresponds to the level of the
Chern-Simons theory. This integer should be odd in the usual
electronic systems since electrons have the fermionic statistics.

When $\nu$ is a more generic fractional number, an important
effective description is given by Chern-Simons theories with
multiple $U(1)$ gauge fields \cite{Wen,Zee}, which is called the
hierarchical (or general) description. Explicitly it is described by
the following action on a three-dimensional spacetime $\Sigma$ (for
details see \cite{Wen}) \be S=\f{1}{4\pi}\int_{\Sigma}\left[
\sum_{i,j=1}^r K_{ij} A^{(i)}\we dA^{(j)} +2 \sum_{i=1}^r q_i
\ti{A}\we dA^{(i)} \right], \label{hir} \ee where $K_{ij}$ is an
integer matrix; $q_i$ are integers and
$\textbf{q}=(q_1,q_2,\dots)$ is called the charge
vector, which characterizes a fractional quantum Hall theory. See
\cite{GMMS} for an analysis of quantum aspects of these theories.
The field $\ti{A}$ is the external $U(1)$ gauge field and the $r$
fields $A^{(i)}$ are dynamical $U(1)^r$ gauge fields that describe
the internal degrees of freedom.

The equations of motion of (\ref{hir}) are
\begin{equation}
K_{ij}\de_{\ap}A^{(j)}_\beta +q_i \de_{\ap}\ti{A}_{\beta}=0.
\end{equation}
The electric current is then given by \be
J^{\ap}\equiv\f{1}{2\pi}\ep^{\ap\beta\gamma}q_i\de_\beta
A^{(i)}_\gamma=\f{1}{2\pi}q_iK^{ij}q_j \ep^{\ap\beta\gamma}\de_\beta
\ti{A}_\gamma, \ee where $K^{ij}$ is the inverse matrix of $K_{ij}$.

Thus we can find the filling fraction
\be
\nu=q_i K^{ij} q_j, \label{fifr}
\ee
and the Hall conductivity
\be
\sigma_{xy}=\f{1}{2\pi}q_i K^{ij} q_j=\f{\nu}{2\pi}.
\ee

A $i$-th quasi-particle ($1\leq i\leq r$) is defined by a particle
which carries a unit charge only with respect to $A^{(i)}$. If we
consider a general quasi-particle  with the charge
$l=(l_1,l_2,\ddd)$, its electric charge becomes \be
Q(l)=l_iK^{ij}q_j, \ee and the phase $e^{i\theta}$ which appears
when we exchange a quasi-particle with the charge $l$ and another
one with $l'$ is given by \be \theta=\pi l_i K^{ij} l'_j. \ee This
angle $\theta$ becomes in general fractional and thus the
quasi-particles obey fractional statistics, called anyons.

In this way the theory is defined by the pair $(K,q)$.
The two theories which are related by
$SL(r,{\bf Z})$ transformation $(K,q)\to (MKM^{T},Mq)$, where $M$
is a $SL(r,{\bf Z})$ matrix, are considered to
be equivalent as they preserve the charge lattice.

One of the most interesting classes of FQH systems can be described
by \be K=[K_{ij}]=\left(
  \begin{array}{cccc}

    a_1 & -1 &  &  \\
    -1 & a_2 & -1 &  \\
     & -1 & a_3 & -1 \\
     &  & -1 &
     \ddots\\
  \end{array}
\right), \ \ \ \ q=(1,0,0,0,\ddd), \label{kmat} \ee where $a_1$
should be an odd integer because electrons have the fermionic
statistics, while $a_{i\geq 2}$ are even integers since all
quasi-particles are bosons, if they are not dressed by statistical
interactions of the Chern-Simons gauge fields. Then the $i$-th
quasi-particle can be regarded as a quasi-particle whose constituent
particles are the $(i-1)$-th quasi-particle. This interesting
structure is called the hierarchy. The filling fraction (\ref{fifr})
is now given by the continued fraction \be
\nu=\f{1}{a_1-\f{1}{a_2-\f{1}{a_3- {\dots}}}}, \ee where the integer
$r$ is the depth of the continued fraction, namely the number of
times we need to take the fractions.

It is a very interesting question as to whether we can realize such
gauge theories in string theory. It would be very hard if we only
consider gauge fields living on D-branes since D-branes usually lead
to non-abelian gauge theories, whose gauge symmetries are not
consistent with the action (\ref{hir}) when $K_{ij}\neq 0$ for
$i\neq j$. Instead we will look for the gauge fields $A^{(i)}$ among
the bulk supergravity fields.

Motivated by the observation
 that the matrix $K$ (\ref{kmat}) resembles the intersection
 matrices of four-dimensional manifolds,
we consider the type IIA string on \be R^{1,2} \times S^3\times
M_4,\label{backg} \ee where we take the four-dimensional manifold
$M_4$ to be an orbifold $\mathbb{C}^2/Z_{n(p)}$. We introduce $k$
units of $H=dB_{NS}$ flux on $S^3$:\footnote{We again set $\al=1$
and follow the convention in \cite{Pol}. } \be \int_{S^3} H=-4\pi^2
k, \ee whose backreaction induces a linear dilaton in one of the
space-like coordinates in $\Sigma=R^{1,2}$ \cite{CHS}. The
background (\ref{backg}) can be realized as the near-horizon limit
of NS 5-branes wrapped on the orbifold $\mathbb{C}^2/Z_n$. We can
add the F-string charge and consider in almost the same way the type
IIA string on $AdS_3\times S^3\times M_4$, which has a clearer
holographic dual via the $AdS_3/CFT_2$ \cite{Maldacena}.

The action of $Z_{n(p)}$ orbifold is defined by $(z_1,z_2)\to
(e^{\f{2\pi i}{n}}z_1,e^{\f{2\pi ip}{n}}z_2)$ where $-n+1\leq p \leq
n-1$ and $(n,|p|)=1$. The orbifold is non-supersymmetric for generic
$p$ and is supersymmetric only when $p=\pm 1$. The type IIA GSO
projection requires $p$ to be an odd integer, otherwise the theory
becomes type 0 string and suffers from a bulk tachyon. Though in
general we cannot avoid localized closed string tachyons
\cite{APS,Localized}, we ignore this issue in this paper.

The minimal resolution of the orbifold singularity
$\mathbb{C}^2/Z_{n(p)}$ is given by a chain of $r$ blown-up 2-cycles
$[i]$ (exceptional $\mathbb{P}^1$ divisors).\footnote{Here
``minimal" means that $r$ is the smallest among all possible
resolutions; namely no 2-cycle can be blown-down without leaving a
singularity.} The intersection number between successive 2-cycles is
$1$ and the self-intersection number of cycle $[i]$ is $-a_i$ where
$a_i\geq 2$ $(i=1,2,\ddd,r)$. The integers $\{a_i\}$ are given by
the Hirzeburch-Jung continued fraction \cite{Ful,Localized}: \be
\f{n}{\tilde{p}}=a_1-\f{1}{a_2-\f{1}{a_3-...}}, \label{fifro} \ee
where $\tilde{p}=p$ for $p>0$ and $\tilde{p}=p+n$ for $p<0$. The
integer $r$ is the depth of the continued fraction. Thus the
intersection matrix of the resolved $\mathbb{C}^2/Z_{n(p)}$ takes
exactly the same form as (\ref{kmat}) except for an overall minus
sign and thus we call it $-\ti{K}_{ij}$.

Now we are interested in the fields in the twisted sectors and such
a theory lives on the fixed points of the orbifold. Then we perform
the Kaluza-Klein compactification on $S^3$ to get a theory on
$R^{1,2}$. We expand the 3-form potential as \be
C_{\mu\nu\rho}=(2\pi)^2\sum_{i}A^{(i)}_{\mu}\omega_{(i)\nu\rho}, \ee
where $\omega_{(i)\mu\nu}$ is the harmonic 2-from dual to the $i$-th 2 cycle,
namely $\int_{[i]}\omega_{(j)}=\delta_{ij}$ and $\ti{K}_{ij}=\int
\omega_{(i)}\we \omega_{(j)}$.

The Chern-Simons term in IIA supergravity leads to \be
\f{1}{4\kappa^2_{10}}\int H\we C\we dC= \f{k}{4\pi}\sum_{i,j=1}^r
\ti{K}_{ij}A^{(i)}\we dA^{(j)}. \ee To couple the
system to an external gauge field,
we add a D4-brane wrapped on a linear combination of 2-cycles
weighted by the charge vector $\textbf{q}$ (with
the resulting 2-cycle denoted by $\sum_{i}q_i[i]$). Then it is easy
to see that the WZ term $\f{1}{(2\pi)^3}\int_{D4}
C \we F$ on this D4-brane gives \be \f{1}{2\pi}\int q_i
\ti{A}\wedge dA^{(i)}, \ee where $\ti{A}$ is the
$U(1)$ gauge field on the D4-brane and serves as the
external gauge field that couples to the FQHE system.

Because the Chern-Simons term dominates over the Maxwell term in the
low-energy theory, the effective theory with the D4-brane coincides
with the Chern-Simons description (\ref{hir}) of the hierarchical
FQHE if we set the minimal value $k=1$. The electric charge is
defined with respect to the gauge field $\ti{A}$. It is also
intriguing to note that the $i$-th quasi-particle (anyon), which has
a unit charge under $A^{(i)}$ via the coupling $\int A^{(i)}$, is
the D2-brane which is wrapped on the 2-cycle $[i]$.

If we go back to the orbifold limit, the wrapped D2
and D4-branes can be regarded as linear combinations
of fractional D0 and D2-branes (for the detailed analysis
refer to \cite{Moore})
. It is very amusing to notice that this fractionality of
D-brane charge is related to
that of FQHE, which essentially comes from the surprising
resemblance between (\ref{fifr}) and (\ref{fifro}).

Thus we have found a remarkable relation between the string theory
on the orbifold and the hierarchical description of the FQHE. It is
intriguing to consider what the holography tells us. In our setup
with F-strings and NS5-branes, we have the IIA string on
$AdS_3\times S^3\times M_4$ so we can apply the $AdS_3/CFT_2$
correspondence. Indeed, the boundary CFT can be identified as the
chiral boson theory which precisely describes the edge state. Note
that the FQH system lives in the bulk $AdS_3$ and the boundary of
the $AdS_3$ is the edge of the FQH system. In this way, we learned
that the edge/bulk correspondence in condensed matter physics, which
is sometimes called `holography' in condensed matter contexts (e.g.\
\cite{FFN}), can be understood as the $AdS_3/CFT_2$ correspondence.

Before we finish this section, we would like to mention the issue of
statistics. One of the most important cases is the supersymmetric
orbifold given by $p=-1$ (ALE singularity). One might think that
this cannot be experimentally realized since it has $a_i=2$ for all
$i$, whereas ordinary FQHE are constructed from strongly-interacting
electrons and fermionic statistics requires $a_1$ to be an odd
integer. Nevertheless, even in realistic FQHEs, it is no longer
absolutely necessary to start from fermionic degrees of freedom.
Indeed, we can realize the FQHE at $\nu=\f{1}{k}$ with $k$ being an
even integer in rotating Bose-Einstein condensates (BECs) in cold
atomic gases \cite{bosonicFQHE}. While cold atoms are neutral under
the electromagnetic $U(1)$ gauge field, rotation plays the role of
magnetic field, as seen from  the Gross-Pitaevskii free-energy
density $\mathcal{F}$ describing a BEC $\phi$ in a frame rotating at
the angular velocity $\vec{\Omega}$, $ \mathcal{F} = \frac{1}{2m}
\left| \left( -i\vec{\nabla} - m (\vec{\Omega}\times \vec{r})
\right)\phi \right|^2 + V(\phi) - \frac{m}{2} (\vec{\Omega}\times
\vec{r}) |\phi|^2 $ \cite{bosonicFQHE}, where $V(\phi)$ is some
potential and $m$ the effective mass of the condensate. If the
angular velocity is large enough and if the filling factor takes a
particular fraction $\nu=1/k$ with $k$ being an even integer, the
cold atoms are expected to be in a bosonic analogue of the Laughlin
FQH state. Similarly, for more general filling fractions which can
be written as continued fractions, the bosonic analogue of the
hierarchical FQH states is expected to be realized, where $a_1$ is
an even integer.

\section{Discussions}

In this paper, we realized three different holographic constructions
of fractional quantum Hall effect
in string theory. Our
models offer us systematic understandings of important theoretical
concepts in FQHE, such as the
fractional quantization of Hall conductivity, edge
states, hierarchy, and topological entanglement entropy. In model I,
the classification of edge states is reduced to that of particular
configurations of D-branes. Similarly in model III, we find that the
classification of hierarchical FQHEs
is equivalent to that of the minimal resolutions of
$\mathbb{C}^2/Z_{n(p)}$ orbifold singularities.
Furthermore, model II reveals a
deep insight holography:
the AdS/CFT correspondence, when applied to the pure $U(N)_k$
Chern-Simons theory, is reduced to the level-rank duality. Since the
main results are already summarized in Section 1, below we would
like to discuss a few
connections between our results and recent developments in FQHE in
condensed matter literature.

In both the first and second models, to apply the AdS/CFT
correspondence, it is necessary to take the large-$N$ limit, whereas
the most common QH systems are described in terms of the $U(1)$ (or
a collection of $U(1)$) Chern-Simons gauge fields. However,
we can break the non-abelian $U(N)$ into $N$ copies of independent
$U(1)$ and extract the FQHE for any one of the $U(1)$ factors as we did in section
2. In section 3, we concentrated on the diagonal $U(1)$ of $U(N)$
and considered FQHE for it. In both models, the holographic
calculations of Hall conductivity seem to be closely related to
those of anomalies.

More interestingly, non-abelian Chern-Simons theories also appear
directly in the effective description of some new types of FQHEs.
For example, $SU(2)$ Chern-Simons gauge theory was suggested to
describe fractional QHE at $\nu=5/2$ \cite{MooreRead, Fradkin97} and
the spin QHE in time-reversal symmetry breaking, singlet topological
superconductors \cite{Read00}. The candidate FQH state for
$\nu=5/2$, called the Moore-Read Pfaffian state, is an example of
non-abaliean FQH states, and excitations build on the ground state
obey non-Abelian statistics. Beside $\nu=5/2$, a FQH state at
$\nu=12/5$ (and $13/5$), which was observed more recently with a
very small energy gap, is also suspected to support non-Abelian
statistics \cite{ReadRezayi}. The prospect of realizing generic
non-abelian Chern-Simons gauge theories in FQHE systems is very
exciting from string theory's perspective: we will have table-top
experiments to realize some of the most fascinating aspects of
string theory, such as holographic duality and D-brane resolution of
orbifold singularity.

We considered the supersymmetric edge states in the ${\cal N}=6$
Chern-Simons theory in model I. They are $\f{1}{2}$ BPS objects and
are described by D4 or D8-branes. The ${\cal N}=6$ Chern-Simons
theory has the $U(N)_k\times U(N)_{-k}$ gauge symmetry. As one way
to relate this system to the realistic FQHE, we simply set $N=1$ and
treat the $U(N)_{-k}$ sector as spectators. We indeed reproduced the
expected result of Hall conductivity. Moreover, if we take this
model literally, our edge states have application beyond the FQHE:
they provide a holographic description for interfaces in any systems
whose effective theory is $\mathcal{N}=6$ Chern-Simons theory. This
is not a remote possibility. In fact, recently it has been suggested
that the double Chern-Simons theory can serve as effective theories
to a number of quantum critical phenomena in realistic materials,
such as the quantum spin Hall effect, the three-dimensional
topological insulators \cite{topinsulators}, and lattice models with
the $Z_2$ topological order \cite{Hansson,WenCS,XS}. It is an
intriguing possibility that string theory can give a nice
classification of general topological insulators in various
dimensions.
\vspace{5mm}

{\bf Note Added:} We noticed that the paper \cite{GaTo}, which
appeared in the arXiv on the same day, has a partial overlap with
this paper in the interpretation of D8-brane charge in the
$AdS_4\times CP^3$ background as the shift of the total level in the
$\mathcal{N}=6$ Chern-Simons theory.

\vskip3mm

\noindent {\bf Acknowledgments}

We are grateful to  M. Fukuma, S. Harashita, S. Hellerman, Y.
Hikida, K. Hori, Y. Imamura, Y. Okawa, H. Ooguri, M. Sato, S. Shiba, S.
Sugimoto, and Y. Tachikawa for useful discussions. This work is
supported by World Premier International Research Center Initiative
(WPI Initiative), MEXT, Japan. The work of TT is also supported in
part by JSPS Grant-in-Aid for Scientific Research No.20740132, and
by JSPS Grant-in-Aid for Creative Scientific Research No. 19GS0219.
SR thanks Center for Condensed Matter Theory at University of
California, Berkeley for its support. WL thanks the high energy
theory group of Kyoto university for their kind hospitality.

\vskip2mm

\appendix

\section{Supersymmetries Preserved by the Edge State Configuration}

\subsection{D4-brane Edge}
\label{D4susy}


Now we show that the D4-brane that wraps on the $\mathbb{CP}^1$ defined by
 $(\theta_1,\vp_1)$ with
$\xi=0$ and that extends along
$r$ and one spatial direction of $AdS_4$ boundary ($R^{1,2}$),
preserves 6 supersymmetries (or 12 supersymmetries by taking into account the
superconformal symmetry). This can be proved in the M-theory lift of ABJM, i.e.
M2-branes probing the orbifold $\mathbb{C}^4/\mathbb{Z}_k$. We take
$(x^0,x^1,x^2)$ to be directions along the M2-branes and
$(x^3,x^4,\ddd,x^{10})$ in the orbifold.

A 11D spinor can be represented as eigenstates: \be
(\gamma_{2},\gamma_{34},\gamma_{56},\gamma_{78},\gamma_{910})\eta
=(s_0,is_1,is_2,is_3,is_4)\eta, \ee where $s_i=\pm 1$. The M2-brane
preserves half of the supersymmetries: $\gamma_{012}\eta=\eta$,
which gives $s_1s_2s_3s_4=1$. Since the orbifold acts as $e^{\f{\pi
i}{k}(s_1+s_2+s_3+s_4)}$, it imposes $\sum_{i}s_i=0$. Namely, the
orbifold projection only allows the following combination:
\begin{equation}\label{s1tos4}
(s_1,s_2,s_3,s_4)=(++--),(+-+-),(+--+),(-+-+),(-++-),(--++),
\end{equation}
(Namely, it projects out $1/4$ of remaining supersymmetries.) Since
each pattern has a degeneracy two (i.e.$s_0=\pm 1$), we have 12
supersymmetries in the ABJM theory.

Now we introduce the M5-brane extending in the
$(x^0,x^1,x^3,x^4,x^5,x^6)$ direction:
\begin{equation}\label{M2-M5}
\begin{array}{r|cccccccccccl}
\,\,  & 0  & 1   & 2   & 3   & 4 & 5   & 6   & 7   & 8   & 9 & 10 \nonumber\\
\,\, \mbox{$R^{1,2}\times \mathbb{C}^4/\mathbb{Z}_k$:}\,\,\, & t  & x
& y   & z_1   & \bar{z}_1 & z_2   & \bar{z}_2   & z_3   & \bar{z}_3   & z_4 & \bar{z}_4 \nonumber\\
\hline N \,\, \mbox{M2:}\,\,\,& \x & \x    &  \x &   &  &   &  & & &
&\,\nonumber\\
1\,\, \mbox{M5:}\,\,\, & \x &  \x &   &  \x & \x & \x  & \x &  & &
&\,
\end{array}
\end{equation}
which preserves the supersymmetry
\begin{equation}
\gamma_{013456}\eta=\eta.
\end{equation}
This means that for each configuration in (\ref{s1tos4}), $\gamma_2$
is now fixed. This breaks half of the supersymmetries: we now have
six supersymmetries. Finally, due to the supersymmetry enhancement
(doubling) at the horizon, the D2-D4 system preserves twelve
supersymmetries.

\subsection{D8-brane Edge}
\label{D8susy}

Now we show that a D8-brane which is wrapped on $\mathbb{CP}^3$ and extends
in the $(t,x,r)$ direction preserves 12 supersymmetries in $AdS_4
\times \mathbb{CP}^3$. The two ten-dimensional 16-components Killing spinors
can be (formally) written as the 11D 32-component Killing spinor
after adding the 11th dimension as $\tilde{y}$ in (\ref{11Dmetric}).
We follow the convention in \cite{NiTa} and set
$(x^0,x^1,\ddd,x^{10})=(t,r,\theta,\phi,\alpha,\beta,\gamma,\xi_1,\xi_2,\xi_3,\xi_4)$.
Then the Killing spinor \cite{NiTa} is given by
\begin{equation}
\epsilon=e^{\frac{\alpha}{2}\hat{\gamma}\gamma_4}
e^{\frac{\beta}{2}\hat{\gamma}\gamma_5}
e^{\frac{\gamma}{2}\hat{\gamma}\gamma_6}
e^{\frac{\xi_1}{2}\gamma_{47}} e^{\frac{\xi_2}{2}\gamma_{58}}
e^{\frac{\xi_3}{2}\gamma_{69}}
e^{\frac{\xi_4}{2}\hat{\gamma}\gamma_{10}}
e^{\frac{\rho}{2}\hat{\gamma}\gamma_{1}}
e^{\frac{t}{2}\hat{\gamma}\gamma_{0}}
e^{\frac{\theta}{2}\gamma_{12}} e^{\frac{\phi}{2}\gamma_{23}}
\epsilon_0,
\end{equation}
where $\epsilon_0$ is a constant 32-component Majorana spinor in
11D.

The
$\Gamma$-matrix that gives the projection operator from D8-brane
is:\footnote{Here we use $\gamma$ to denote tangent space gamma
matrices, and $\Gamma$ space-time ones.} \ba
\Gamma_{D8}&=&\gamma_{01}\Gamma_x\gamma_{\sigma_1\cdots\sigma_6}\gamma_{10},\\
\Gamma_x&=&\frac{1}{\sqrt{(\partial_x\theta)^2+\sin^2\theta(\partial_x\phi)^2}}
(\partial_x\theta\gamma_2+\partial_x\phi\gamma_3). \ea It can be
greatly simplified:
\begin{equation}\label{D8susycondition}
\Gamma_{D8}=
\frac{1}{\sqrt{(\partial_x\theta)^2+\sin^2\theta(\partial_x\phi)^2}}
(\partial_x\theta\gamma_3-\partial_x\phi\gamma_2)\equiv \Gamma_y.
\end{equation} Here $y$ is the orthogonal direction of $x$. D8-brane preserves
only those parts of the Killing spinors which satisfy
\begin{equation}
\Gamma_{D8}\epsilon=\epsilon.
\end{equation}
There are two solutions, which are equivalent by a rotation:
\begin{eqnarray} &&\theta=x,\qquad
\phi=\phi_0, \qquad\qquad \textrm{with}\quad(-\sin\phi_0 \gamma_2
+\cos \phi_0
\gamma_3)\epsilon_0=\epsilon_0\\
\textrm{or}\quad&&\theta=\frac{\pi}{2},\qquad \phi=x, \qquad\qquad
\textrm{with}\quad \gamma_1 \epsilon_0=-\epsilon_0.
\end{eqnarray}
Now we can repeat the argument for D2-D4 system. Instead of
$\gamma_2$, we simply choose $\epsilon_0$ to be the eigenstate of
$-\sin\phi_0 \gamma_2 +\cos \phi_0 \gamma_3$ for the first solution
or $\gamma_1$ for the second one. Then condition
(\ref{D8susycondition}) breaks half of 24 supersymmetries
preserved in $AdS_4 \times \mathbb{CP}^3$. Thus the D8-brane preserve 12
supersymmetries.

\section{Solution of the Bending Probe Brane at Zero Temperature}
\label{BendingBraneExactSol}

Here we would like to present an exact solution to (\ref{eomymm}) at
zero temperature. First it is useful to notice that  we
have the 'Hamiltonian' \be
D=1+z'^2+z^4F_{xy}^2-z^4F_{ty}^2-z^4(1+z'^2)F_{tx}^2=\f{d^2}{z^6},
\ee where $d>0$ is a constant.

After we perform a constant shift of $f(y)$ and $g(y)$ (called
$F(y)$ and $G(y)$), we find the equations of motion are conveniently
summarized as follows: \ba
&& F'=-\kappa\f{\s{D}}{z}G,\ \ \ \ \ \ G'=-\kappa\f{\s{D}}{z}F,\label{csf}\\
&& (z')^2=\f{d^2}{z^6}(1-\kappa^2\beta^2 z^2)-1. \label{dbp} \ea
where we set $\kappa=\f{\ti{k}}{2\pi\ap}$; $\beta$ is also a
constant defined by \be F^2-G^2=\beta^2. \ee

The profile of the D-brane is determined by solving (\ref{dbp}). Its
turning point is $z=z_*$ defined by
$\f{d^2}{z^6_*}(1-\kappa^2\beta^2 z^2_*)-1=0$.

We can solve $F$ and $G$ as \be F+G=A_1\cdot e^{I(y)},\ \ \ \
F-G=A_2\cdot e^{-I(y)}, \ee where $A_1A_2=\beta^2$ and the function
$I(y)$ is \be I(y)=\int^{z(y)}_{z_*} dz
\f{kd}{z\s{d^2(1-\kappa^2\beta^2 z^2)-z^2}}. \ee

If we denote an odd and even functions which are two independent
solutions for $F$ by $F_1(y)$ and $F_2(y)$, then we can write
\be\label{BBAtAxsolT0} A_t(y)=a_1F_1(y)+a_2F_2(y)+a_3,\ \ \ \
A_x(y)=a_1F_2(y)+a_2F_1(y)+a_4, \ee where $a_1, a_2, a_3$ and $a_4$
are all integration constants.

\section{Solution for Single-Edge at Zero Temperature}
\label{SingleBraneExactSol}

If we assume the zero temperature i.e. $r_0=0$,
the profile of $A_x=g(z)$ and $A_t=f(z)$ can be solved
exactly. Their EOMs become
\be
\f{zF'}{1-z^4(F')^2+z^4(G')^4}=-\f{\ti{k}}{2\pi\ap}G,\ \ \ \
\f{zG'}{1-z^4(F')^2+z^4(G')^4}=-\f{\ti{k}}{2\pi\ap}F,
\ee
where we defined
\be
F=f+\f{2\pi}{\ti{k}}j-\f{2\pi}{\ti{k}}\left(\f{\ti{k}}{4\pi}+\eta\right)f(0),\ \ \ \ G=g-\f{2\pi}{\ti{k}}\left(\f{\ti{k}}{4\pi}-\eta\right)g(0)-\f{2\pi}{\ti{k}}\rho.
\ee

Below we assume $\ti{k}<0$ and define $\kappa=\f{|\ti{k}|}{2\pi\ap}$.

First we notice $(F^2-G^2)'=0$. Thus we can set
\be
G^2=F^2+\beta^2,
\ee
where $\beta$ is a constant. The relation $G^2>F^2$ will be made clear later.

Then the EOMs are reduced to the differential equation for $F(z)$
\be
F'=\f{\kappa\s{F^2+\beta^2}}{z\s{1+\kappa^2\beta^2 z^2}}.
\ee

Its solution is given by

\be
F+\s{F^2+\beta^2}=A\left(\f{z}{1+\s{1+\kappa^2\beta^2 z^2}}\right)^{\kappa}.
\ee

In summary we find
\ba
2F=A\left(\f{z}{1+\s{1+\kappa^2\beta^2 z^2}}\right)^{\kappa}-\f{\beta^2}{A}\left(\f{z}{1+\s{1+\kappa^2\beta^2 z^2}}\right)^{-\kappa}, \no
2G=A\left(\f{z}{1+\s{1+\kappa^2\beta^2 z^2}}\right)^{\kappa}+\f{\beta^2}{A}\left(\f{z}{1+\s{1+\kappa^2\beta^2 z^2}}\right)^{-\kappa}.
\ea

Thus the general solutions take the form
 \ba
F=-\f{\beta^2}{A}z^{-\kappa}+Az^\kappa +\ddd,\ \  \no
G=\f{\beta^2}{A}z^{-\kappa}+Az^\kappa +\ddd.\ \  \label{nort}
\ea
The leading term $z^{-\kappa}$ implies that the operator
 to the gauge field has the conformal dimension
$\Delta=1+\kappa~(>1)$ \cite{Mi}.

\section{Another Derivation of Hall Conductivity in D3-D7 Model}
\label{HallD3D7}

Here we would like to compute the Hall conductivity in the
holographic dual gravity side treating the D7-branes as probe
D-branes instead of replacing it with the background axion flux.
Since we are interested in the Hall conductivity in the low energy
theory, we place the holographic screen (i.e. the boundary of AdS)
near $r=r_0$: at $r=r_1$ with $r_1\gtrsim r_0$. Then the holographic
current is computed as the on-shell variation of the action:\be
j^\mu=\f{\delta S_{IIB}}{\delta A_{RR\mu}}|_{r=r_1}, \ee where
$A_{RR}$ is the 1-form that comes from the RR 2-form field $B_{RR}$
via (\ref{brr}) and serves as the external gauge field in the
D3-brane system. $S_{IIB}$ is the type IIB supergravity action plus
the D7-brane world-volume action evaluated on-shell. In the
leading-order calculation of the holographic current, only $B_{RR}$
and $B_{NS}$ are important; and the type IIB supergravity action
includes their kinetic terms: \be -\f{1}{4\kappa^2}\int
d^{10}x\sqrt{-g}\left(g_s^{-2}|H_{NS}|^2+|H_{RR}|^2\right)
=-\f{1}{2(2\pi)^7}\int
d^{10}x\sqrt{-g}\left(g_s^{-2}|H_{NS}|^2+|H_{RR}|^2\right) . \ee In
the presence of the 5-form flux, the Chern-Simons term
$-\f{1}{4\kappa^2}\int C_4\we H_{NS}\we H_{RR}$ together with the
cross terms in $-\f{1}{8\kappa^2}\int|\ti{F}_{5}|^2$ gives a
Chern-Simons coupling (after compactifying on $S^5$) \cite{AW} \be
\f{N}{(2\pi)^3}\int dB_{NS}\we B_{RR}. \label{csrn} \ee We fixed the
surface term in (\ref{csrn}) by requiring that the gauge symmetries
$B_{NS}\to B_{NS}+d\Lambda_{NS}$ and $B_{RR}\to
B_{RR}+d\Lambda_{RR}$ are preserved even when $H_{RR}\neq 0$.

To write down the D7-brane world-volume action, we
will use the ansatz in which the $U(k)$ gauge field $A$ is
proportional to the identity matrix $\textbf{1}_{k\times k}$. Then
terms involving $A$ are \be \f{N}{4\pi}\int_{R^{1,2}} \mbox{Tr}
A\we dA +\f{N}{(2\pi)^2}\int_{R^{1,2}} \mbox{Tr} A\we B_{NS}+
\f{1}{(2\pi)^6}\int_{D7}\mbox{Tr}A\we *dB_{RR}, \label{sevena} \ee
where in the first two terms
we have integrated the RR 5-form flux over $S^5$
and the last term comes from the RR-coupling $2\pi T_{7}\int
\mbox{Tr}F\we C_{(6)}$.

The equation of motion of the gauge field $A$ on D7-branes integrated
over $S^5$ leads to \be \f{N}{2\pi}dA+\f{1}{(2\pi)^6}\int_{S^5}
\left(*H_{RR}\right)|_{D7}=0, \label{dseoma} \ee where $\ddd |_{D7}$
means the restriction to (pullback to) the D7-brane
world-volume.

Now we assume
a Dirichlet boundary condition for $B_{RR}$ and a Neumann one for
$B_{NS}$ at the boundary $r=r_1$. Also we expect that
the low-energy solution has $B_{NS}=0$.  The variation of $B_{NS}$ vanishes
 at $r=r_0$ by a
cancellation between the bulk and D7-branes contributions only if \be
A_{RR}|_{D7}+kA=0,  \label{dseomb} \ee using (\ref{csrn}),
(\ref{sevena}) and (\ref{brr}). Note that the two equations
(\ref{dseoma}) and (\ref{dseomb}) determine $A_{RR}$ completely.

We are ready to compute the holographic charge density $\rho$ and
current density $j$. Define a 1-form $J\equiv\rho dt+j_x dx+j_y dy$
and assume $r_1$ is very close to $r_0$:
 \be J=
\f{\delta S_{IIB}}{\delta A_{RR}}\Bigr |_{r=r_1}=\f{1}{(2\pi)^7}
*_3\int_{S^1\times S^5} d\ti{\theta}\we (*
H_{RR})|_{D7} =\f{N}{2\pi} *_3(dA)=\f{N}{2 \pi k}
*_3(F_{RR}|_{D7}). \label{J} \ee
where the first step is an on-shell variation of the
kinetic term $\int_{M} d(\delta B_{RR})\we *H_{RR}$ which leads to
the boundary variation $\int_{\de M} \delta B_{RR}\we *H_{RR}$ after
imposing the bulk equation of motion of $B_{RR}$; and we used
(\ref{dseoma}) and (\ref{dseomb}) in the last two steps.

The $x$-component of (\ref{J}) gives $j_x=\f{N}{ kh}E_y$. Thus we
find the fractional Hall conductivity \be \sigma_{xy}=\f{N}{kh}. \ee
This clearly agrees with what we find from (\ref{HCS}).






\end{document}